\documentclass[nohyperref]{article}

\usepackage{microtype}
\usepackage{graphicx}
\usepackage{subcaption}
\usepackage{booktabs} 
\usepackage{wrapfig, cutwin}
\usepackage{enumerate}
\usepackage{pgfplots}
\pgfplotsset{compat=1.18}
\usepackage{comment}
\usepackage{placeins}

\usepackage{hyperref}

\definecolor{mitred}{rgb}{0.6,0.2,0.2}
\definecolor{mitgray}{rgb}{0.4,0.4,0.4}
\definecolor{darkgray}{cmyk}{0,0,0,0.90}
\definecolor{lightgray}{cmyk}{0,0,0,0.50}

\usepackage[accepted]{icml2024}

\usepackage{amsmath}
\usepackage{amssymb}
\usepackage{mathtools}
\usepackage{amsthm}
\usepackage{bm, dsfont}
\usepackage[normalem]{ulem}

\renewcommand{\rm}[1]{\mathrm{#1}}

\newcommand{\ind}{\mathds{1}}

\newcommand{\RR}{\mathbb{R}}

\usepackage[capitalize,noabbrev]{cleveref}

\theoremstyle{plain}
\newtheorem{theorem}{Theorem}[section]
\newtheorem{proposition}[theorem]{Proposition}
\newtheorem{lemma}[theorem]{Lemma}
\newtheorem{corollary}[theorem]{Corollary}
\theoremstyle{definition}

\newtheorem{assumption}[theorem]{Assumption}
\theoremstyle{remark}
\newtheorem{remarkx}[theorem]{Remark}
\newenvironment{remark}
  {\pushQED{\qed}\remarkx}
  {\popQED\endremarkx}
\newtheorem{examplex}[theorem]{Example}
\newenvironment{example}
  {\pushQED{\qed}\examplex}
  {\popQED\endexamplex}

\usepackage[textsize=tiny]{todonotes}

\icmltitlerunning{Multiply-Robust Causal Change Attribution}

\begin{document}
\usetikzlibrary{arrows.meta}
\twocolumn[
\icmltitle{Multiply-Robust Causal Change Attribution}




\begin{icmlauthorlist}
\icmlauthor{Victor Quintas-Martinez}{mmm,xxx}
\icmlauthor{Mohammad Taha Bahadori}{aaa}
\icmlauthor{Eduardo Santiago}{aaa}
\\ \icmlauthor{Jeff Mu}{aaa}
\icmlauthor{Dominik Janzing}{aaa}
\icmlauthor{David Heckerman}{aaa}
\end{icmlauthorlist}

\icmlaffiliation{aaa}{Amazon}
\icmlaffiliation{mmm}{MIT Department of Economics.}
\icmlaffiliation{xxx}{This work was completed while VQM was an intern at Amazon.}

\icmlcorrespondingauthor{Victor Quintas-Martinez}{vquintas@mit.edu}

\icmlkeywords{Machine Learning, ICML}
\vskip 0.2in
]



\printAffiliationsAndNotice{}  

\begin{abstract}
Comparing two samples of data, we observe a change in the distribution of an outcome variable. In the presence of multiple explanatory variables, how much of the change can be explained by each possible cause? We develop a new estimation strategy that, given a causal model, combines regression and re-weighting methods to quantify the contribution of each causal mechanism. Our proposed methodology is multiply robust, meaning that it still recovers the target parameter under partial misspecification. We prove that our estimator is consistent and asymptotically normal. Moreover, it can be incorporated into existing frameworks for causal attribution, such as Shapley values, which will inherit the consistency and large-sample distribution properties. Our method demonstrates excellent performance in Monte Carlo simulations, and we show its usefulness in an empirical application. Our method is implemented as part of the Python library \texttt{DoWhy} \citep{dowhy,dowhy_gcm}.
\end{abstract}

\section{Introduction}\label{sec:intro}
Analysts are often interested in identifying and quantifying the contribution of multiple possible causes of change to the performance metrics of large-scale systems, such as sales volume, throughput, or retention rate.
As an example, a manufacturer compares data from 2023 and 2022 and realizes that net sales have increased. However, many factors have changed between these two time periods, including the characteristics of the product, competitors' prices, or market conditions. How much did each of these variables contribute to the increase in average sales? This is a \emph{change attribution} question. When policymakers and business leaders are interested in using these insights for future changes, the change attribution problem necessarily becomes \emph{causal}.

In the \textit{causal change attribution} problem, we observe two samples of data, including an outcome and multiple explanatory variables. We are also given a causal Directed Acyclic Graph (DAG) that encodes the prior expert knowledge about the causal conditional (in-)dependence relationships among the variables in the data. The objective is to assign a score to each of the causal mechanisms in this DAG that quantifies its contribution to the change in the distribution of the outcome. Although in many instances we observe a change over time, we want to emphasize that our methods apply to more general settings involving a comparison of two samples --- for example, differences between groups (e.g., females and males) or geographical locations (e.g., East vs. West Coast of the US). 

The contribution of each causal mechanism is generally very difficult to disentangle. We need to characterize what the distribution of the outcome variable would have been if we shifted only some causal mechanisms, but left others unchanged. This is a \textit{counterfactual} distribution, which is not directly observable in the data. We overcome this fundamental challenge by employing a combination of \textit{regression} and \textit{re-weighting} methods. Moreover, our estimating equation is multipy robust, in the sense that it still recovers the parameters of interest even if some components of the model are misspecified. We show that our estimator is consistent and asymptotically normal under weak conditions on the ML algorithms used to learn the regression and the weights. The asymptotic variance is also consistently estimable, allowing us to compute valid standard errors, confidence intervals and $p$-values. To obtain these results we build on the existing double/debiased ML literature \citep{chernozhukov2018double, chernozhukov2021automatic, chernozhukov2023automatic}.

There is a large body of work on the attribution problem \citep{efron2020prediction, yamamoto2012understanding, dalessandro2012causally}, some of which has studied \textit{causal} attribution \citep{liu2023need, lu2023evaluating, mougan2023explanation, zhao2023conditional, berman2018beyond, ji2016probabilistic, dawid2014fitting, shao2011data}. An important branch of the literature has focused on developing suitable definitions of the contribution of each variable, e.g., with variations of Shapley values \citep{janzing2024quantifying, chen2023algorithms, budhathoki2022causal, jung2022measuring, sharma2022counterfactual} or optimal transport methods \citep{kulinski2023towards}, while remaining agnostic about estimation. Our contribution is related but distinct: the estimator we develop can be readily integrated into existing frameworks for causal change attribution, such as Shapley values, 
or it may be of interest in its own right. Shapley values based on our estimator will inherit its consistency and asymptotic normality properties; we also propose a convenient and computationally efficient bootstrap procedure to compute their standard errors.

Although other recent algorithms have been proposed to estimate Shapley Values, a majority of these algorithms are not multiply-robust, and generally they are based on regression only (learning the distribution of the outcome given the explanatory variables).\footnote{The only exception that we are aware of is \citet{jung2022measuring}. However, their notion of $do$-Shapley values decomposes the effect of fixing a value for the explanatory variables, rather than a distribution for each causal mechanism, as we do in this paper. They consider only discrete-valued explanatory variables, and their approach does not seem easily generalizable beyond that.} As such, they will not perform well unless we have a high-quality estimator of the regression function. In high-dimensional or non-parametric settings, estimators for those objects will typically exhibit slow convergence rates, rendering traditional normal-based large sample inference (standard errors, confidence intervals or $p$-values) invalid. We provide the first formal analysis of the asymptotic properties of an estimator in the causal change attribution setting of \citet{budhathoki2021did}, which allows us to perform valid inference in large samples (standard errors, confidence intervals and tests).

Causal change attribution is related to the classical problem of treatment effect estimation \citep{pearl2009causality,imbens2015causal,peters2017elements} in the need to evaluate counterfactual distributions, but there are some fundamental differences. 
Within the treatment effects literature, the closest problem to ours is that of \textit{causal mediation}. \citet{tchetgen2012semiparametric} gave multiple-robustness results for mediation analysis with a single mediator (explanatory variable). The case with multiple mediators and an arbitrary DAG is substantially more challenging. Although causal mediation with multiple mediators has been studied before (e.g., \citealp{daniel2015causal}), we provide the first multiply-robust estimator in this setting. We also believe to be the first to explicitly build a connection between the causal change attribution problem and the causal mediation literature.

\section{Methodology}\label{sec:theory}
In this section we describe our multiply-robust methodology for causal change attribution. \cref{sec:setting} defines the causal model and the causal change attribution problem and its connections and contrasts with treatment effect estimation. \cref{sec:examp} gives an identification result in a simple example, which we then generalize in \cref{sec:ident}. \cref{sec:est} describe how to implement our estimator, and \ref{sec:inference} gives large-sample inference results. Finally, in \cref{sec:combin} we explain how our method can be incorporated within existing frameworks of causal change attribution, such as Shapley values. We provide the technical conditions and proofs for all lemmas and theorems in Appendix \ref{sec:proofs}.

\subsection{Setting} \label{sec:setting}
We observe multiple i.i.d. measurements of the same variables $(T, \bm{X}, Y)$. Each observation consists of a sample indicator $T \in \{0, 1\}$,  $K$ explanatory variables $\bm{X} := (X_1, \ldots, X_K)$ and an outcome of interest $Y \in \mathcal{Y} \subset \RR$. The explanatory variables in $\bm{X}$ could be continuous, categorical, or even unstructured data types such as text or images.

We assume that the distribution of $(\bm{X}, Y) \mid T = t$ has a causal Markov factorization \citep{spirtes2000causation}:
\begin{equation}
    P_{(\bm{X}, Y)}^{(t)} := P_{Y \mid \bm{X}}^{(t)} \prod_{k=1}^{K} P_{X_k \mid \rm{PA}_k}^{(t)}, \tag{F} \label{eq:factor}
\end{equation}
where $\rm{PA}_k$ are the \emph{parents} (direct causes) of $X_k$ in the underlying causal DAG, other than $T$. Throughout, the superscript $(t)$ denotes conditioning on $T = t \in \{0, 1\}$. Each conditional distribution on the right-hand side of \eqref{eq:factor} is called a \textit{causal mechanism} \citep{peters2017elements}. Without loss of generality, we assume that the explanatory variables are labeled so that $k < k'$ if $X_k \in \rm{PA}_{k'}$, i.e., the causal antecedents of $X_{k'}$ in the DAG have indices lower than $k'$. In practice, the researcher only needs to know one such \textit{causal ordering}, rather than the full causal graph for $(\bm{X}, Y)$. Our causal model implies a distribution for $Y \mid T = t$ by marginalization, i.e., \[P_{Y}^{(t)} := \int P_{Y \mid \bm{X}}^{(t)} \prod_{k=1}^{K} \rm{d}P_{X_k \mid \rm{PA}_k}^{(t)}.\] The problem of \textit{change attribution} tries to quantify how much of the differences between $P_{Y}^{(1)}$ and $P_{Y}^{(0)}$ are due to shifting each causal mechanism from its distribution at $T = 0$ to its distribution at $T = 1$. To clarify, intervening on a causal mechanism $P_{X_k \mid \rm{PA}_k}$ may impact the outcome both directly and indirectly through other explanatory variables which are causal descendants of $X_k$. We are interested on the total effect of changing $P_{X_k \mid \rm{PA}_k}$, without fixing the marginal distribution of its causal descendants. Finally, note that we allow the causal mechanisms to differ between samples, but the underlying DAG is assumed to be the same.

\begin{figure}[!t]
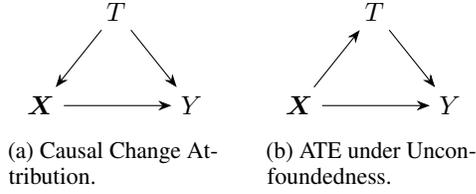

    \centering
    \subfloat[Causal Change Attribution. \label{fig:dagsattr}]{\input{figures/attr}} \qquad
    \subfloat[ATE under Unconfoundedness. \label{fig:dagsate}]{\input{figures/ate}}
    \caption{Two possible DAGs for $(T, \bm{X}, Y)$.}
    \label{fig:dags}
\end{figure}

We want to draw a clear distinction between attribution and a related problem that has received a much larger attention in the literature, Average Treatment Effect (ATE) estimation under unconfoundedness. We depict the difference between the causal models underlying these two problems in \cref{fig:dags}. In an ATE setting, $\bm{X}$ typically represents pre-treatment covariates, that affect both the outcome $Y$ and the propensity to receive a binary treatment $T$. To compute the causal effect of $T$ we need to control for those covariates, that is, compare units with similar values of $\bm{X}$. In contrast, causal change attribution acknowledges that both the distributions of $\bm{X}$ and of $Y$ given $\bm{X}$ vary depending on whether $T = 0$ or $T = 1$. The goal of causal change attribution is to quantify how much of the effect of $T$ on $Y$ goes through (is mediated by) each causal mechanism in this distribution. Under an unconfoundedness assumption about the assignment of $T$, the problem of causal attribution can be thought of as decomposing the ATE of $T$ on $Y$ into the Natural Direct Effect and the Natural Indirect Effect corresponding to each causal mechanism (\citealp[\S4.5]{pearl2009causality}; \citealp{daniel2015causal}). We describe this in more detail, including the corresponding structural equation model and potential outcomes, in \cref{sec:structural}.

We note, however, that our definition of attribution does not require $T$ to be a treatment that can be administered in the ``interventional'' sense. As discussed, $T$ could also be an indicator for group membership (e.g., female or male), time period (e.g., before and after a certain date), or geographical location (e.g., East vs. West Coast of the US). Under which conditions can this change attribution be interpreted as \textit{causal}? We will assume that the researcher knows a causal ordering of the underlying true causal graph for $(\bm{X}, Y)$. This will allow us to quantify the contribution of each \textit{causal mechanism}, rather than changes in the \textit{marginal distributions} of explanatory variables. In practice, researchers can use a combination of causal discovery methods (e.g., the conditional independence test of \citealp{zhang2011kernel}) and domain knowledge to obtain the DAG. Finally, we also assume that there is no unobserved variable $U$ such that $T \rightarrow U$, $U \rightarrow \bm{X}$ and $U \rightarrow Y$. We discuss ways to relax this assumption in future research in \cref{sec:conclusions}.

\subsection{Preliminary Example}\label{sec:examp}
We begin with a simple yet illustrative example. Consider the simplest situation of $K = 1$. How much of the difference in means $\rm{E}[Y \mid T = 1] - \rm{E}[Y \mid T = 0]$ is due to changing $P_{X}$? How much of it is due to changing $P_{Y \mid X}$?

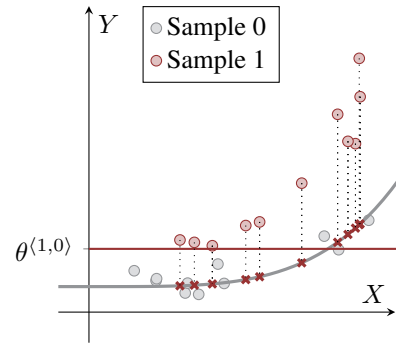
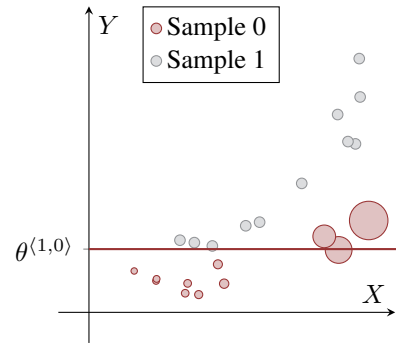
\begin{figure}[!b]
    \centering
    \subfloat[Regression: We learn the dependence between $Y$ and $X$ in sample 0 through the regression function $\gamma(X)$ (gray curve). Subsequently, we average the fitted values for $X$ in sample 1 (red crosses).]{\qquad\begin{tikzpicture}
\begin{axis}[height=2\columnwidth,
   width=2\columnwidth, scale = 0.3,
   clip mode=individual,
   axis x line=center,
   axis y line=center,
   xmin = -0.1, xmax = 1,
   ymin = -0.3, ymax = 3,
   xlabel=$X$,ylabel=$Y$,
   ytick=0.62,xtick=\empty,
   legend style={at={(0.25, 1.0)},anchor=north west},
   xticklabel=\empty,
   yticklabel={$\theta^{\langle 1,0 \rangle}$}]
\addplot [only marks, mark options={fill=lightgray!30, draw=lightgray}] table [x="x0", y="y0", col sep=comma] {figures/data.csv};
\addlegendentry{Sample 0};
\addplot [only marks, mark options={fill=mitred!30, draw=mitred}] table [x="x1", y="y1", col sep=comma, meta index=1] {figures/data.csv}
\foreach \i in {0,...,20} {
                coordinate [pos=\i/20] (a\i)
            };
\addlegendentry{Sample 1};
\addplot [very thick, draw=lightgray, samples=50, domain=-0.1:1.2] {x^4 + 0.25};
\addplot [only marks, mark=x, mark options={draw=mitred}, very thick] table [x="x1", y="r0", col sep=comma, meta index=1] {figures/data.csv}
\foreach \i in {0,...,20} {
                coordinate [pos=\i/20] (b\i)
            };
\addplot[mitred, thick, domain=0:1.2] {0.62};
\end{axis}
\foreach \i in {0,...,20} {
            \draw [dotted] (a\i) -- (b\i);
        }
\end{tikzpicture}\qquad\qquad}
    
    \vspace{1em}
    \subfloat[Re-weighting: We average $Y$ in sample 0 (red circles), but we give more weight to observations whose $X$ is more likely to come from sample 1.]{\qquad\begin{tikzpicture}
\begin{axis}[height=2\columnwidth,
   width=2\columnwidth, scale = 0.3,
   clip mode=individual,
   axis x line=center,
   axis y line=center,
   xmin = -0.1, xmax = 1,
   ymin = -0.3, ymax = 3,
   xlabel=$X$,ylabel=$Y$,
   ytick=0.62,xtick=\empty,
   legend style={at={(0.25, 1.0)},anchor=north west},
   xticklabel=\empty,
   yticklabel={{$\theta^{\langle 1,0 \rangle}$}}
   ]
\addplot [scatter, only marks, color=mitred!30, draw=mitred,
    point meta=explicit symbolic,
    scatter/@pre marker code/.style={/tikz/mark size=1+\pgfplotspointmeta},
    scatter/@post marker code/.style={}
    ] table [x="x0", y="y0", meta="dr", col sep=comma] {figures/data.csv};
\addlegendentry{Sample 0};
\addplot [only marks, mark options={fill=lightgray!30, draw=lightgray}] table [x="x1", y="y1", col sep=comma] {figures/data.csv};
\addlegendentry{Sample 1};
\addplot[mitred, thick, domain=0:1.2] {0.62};
\end{axis}
\end{tikzpicture}\qquad\qquad}
    \caption{Visual intuition for regression and re-weighting.}
    \label{fig:visual}
\end{figure}

In order to answer these questions, we would like to know what the mean of $Y$ would be if we shifted $P_{X}$ to be as in sample 1, but left $P_{Y \mid X}$ unchanged as in sample 0, which we denote as: 
\[\theta^{\langle 1, 0 \rangle} = \int y \rm{d}P^{\langle 1,0 \rangle}_{Y}(y), \quad \text{for} \quad P^{\langle 1,0 \rangle}_{Y} = \int P_{Y\mid X}^{(0)} \rm{d} P_{X}^{(1)}. \] 
The fundamental challenge is that $P^{\langle 1,0 \rangle}_{Y}$ is a \textit{counterfactual} distribution \citep[\S4.5]{pearl2009causality}, in the sense that we don't observe data sampled from it, and so we cannot estimate $\theta^{\langle1,0\rangle}$ directly as a sample average. It is still possible, however, to identify $\theta^{\langle1,0\rangle}$, as shown in the following lemma:
\begin{lemma}\label{lem:identexamp}
Under the regularity conditions given in the appendix, we have the following identification results:
\begin{align}
    \theta^{\langle 1,0 \rangle} & = \rm{E}_{(1)}[\gamma(X)] \tag{REG} \label{eq:reg} \\
     & = \rm{E}_{(0)}[\alpha(X) Y] \tag{REW} \label{eq:rew},
    \end{align} 
where $\gamma(X) := \rm{E}_{(0)}[Y \mid X]$,  $\alpha(X) := {\rm{d}P_{X}^{(1)}}/{\rm{d}P_{X}^{(0)}}(X)$,
and $\rm{E}_{(t)}[\cdot]$ denotes the expectation conditional on $T = t$.
\end{lemma}

We show the intuition for \cref{lem:identexamp} graphically in \cref{fig:visual}. Equation \eqref{eq:reg} gives identification by \emph{regression}. We learn the dependence between $Y$ and $X$ in sample 0 through a (non-parametric) regression, and then average that regression function over the $X$ in sample 1. This is essentially a non-parametric generalization of the Oaxaca-Blinder decomposition \citep{oaxaca1973male, blinder1973wage}. It also appears in the causality literature as part of the mediation formula (see \cref{sec:structural} and \citealp[\S4.5]{pearl2009causality}). Equation \eqref{eq:rew} gives identification by \emph{re-weighting}. We average $Y$ in sample 0, but we give more weight to observations whose $X$ is more likely to come from sample 1. Formally, the weights $\alpha(X)$ are Radon-Nykodim (RN) derivatives \citep{billingsley1995probability}. When $X$ has a density or a probability mass function, these are simply the ratio of densities or probability mass functions between the two samples, respectively. The re-weighting idea has been used in the literature on covariate shift problems \citep{shimodaira2000improving, bickel2009discriminative}, but we are not aware of causal change attribution methods that leverage this insight.

\begin{remark}[Relation to Mediation]
Causal Change Attribution is closely related to decomposing the total effect of an intervention that sets $T = 1$ into the Natural Direct Effect and the Natural Indirect Effect of \citet[\S4.5]{pearl2009causality}. We discuss this connection in detail in \cref{sec:structural}.
\end{remark}

The following result combines regression and re-weighting methods to obtain a more robust identifying equation. It is essentially a version without pre-treatment covariates of the efficient influence function given in \citet{tchetgen2012semiparametric} for causal mediation analysis with a single mediator. We restate it here in our notation, because it will make the intuition of our novel result for a general causal graph (\cref{sec:ident}) clearer.
\begin{lemma}\label{lem:exampleDR} Let $g(X)$, $a(X)$ be two functions 
such that $\rm{E}_{(0)}[g(X)^2] < \infty$, $\rm{E}_{(0)}[a(X)^2] < \infty$.
Consider the following estimating equation:
\begin{equation}
  \rm{E}_{(1)}[g(X)] + \rm{E}_{(0)}[a(X) e(X, Y)],  \tag{DR} \label{eq:DRexamp}
\end{equation}
where $e(X, Y) := Y - g(X)$.

Under the conditions of Lemma \ref{lem:identexamp}, \eqref{eq:DRexamp} 
is equal to $\theta^{\langle 1,0 \rangle}$ if $g(X) = \gamma(X)$ or $a(X) = \alpha(X)$, but not necessarily both.
\end{lemma}

Equation \eqref{eq:DRexamp} is \emph{doubly robust} in the following sense: it still identifies the parameter of interest even if one of the regression function or the weights is misspecified. The term $\rm{E}_{(0)}[a(X) e(X, Y)]$ is a ``debiasing'' term, as in the double/debiased machine learning literature \citep{chernozhukov2018double}, consisting of an average of the non-parametric regression error $e(Y, X)$ weighted by $a(X)$. If the regression function is correctly specified, this average will be zero regardless of the weights (by the Law of Iterated Expectations). On the other hand, if the regression function is not correctly specified but the weights are, the second term will account and correct for the misspecification of $g(X)$. We refer the reader to the proof in \cref{sec:proofs} for details.

\begin{remark}[On the Overlap Assumption]\label{rem:overl}
     One of the regularity conditions (\cref{ass:overl}) imposes that the support of $P^{(0)}_{X}$ includes the support of $P^{(1)}_{X}$. From a technical perspective, this assumption guarantees that the RN derivative $\alpha(X)$ exists. In this remark, we give a more intuitive explanation. The regression strategy \eqref{eq:reg} requires estimating a regression of $h(Y)$ on $X$ in sample 0, and then obtaining the fitted values in sample 1. Without overlap, we would be extrapolating. In general, we want to avoid this (unless we have a credibly good parametric model for the regression). Similarly, the re-weighting strategy \eqref{eq:rew} breaks down without overlap, because some regions of $X$ values that have positive probability in sample 1 are never observed in sample 0. For our asymptotic results in \cref{sec:inference} we will impose a stronger form of overlap (\cref{ass:stroverl}), which is analogous to the overlap assumption in ATE estimation under unconfoundedness.
\end{remark}

\subsection{Identification}\label{sec:ident}
We are now ready to introduce the main result. Let $\bm{c} := \langle c_1, \ldots, c_K, c_{K+1} \rangle \in \{0,1\}^{K+1}$ denote a \emph{change vector}, where $c_k = 1$ if we shift the $k$-th causal mechanism to be as in sample 1, and $c_k = 0$ otherwise. The last entry $c_{K+1}$ indicates whether we want to shift the conditional distribution of the outcome, $P_{Y \mid \bm{X}}$. We will denote by $P^{\bm{c}}_{Y}$ the distribution:
\[P^{\bm{c}}_{Y} := \int P_{Y \mid \bm X}^{(c_{K+1})}\prod_{k=1}^K \rm{d}P^{(c_k)}_{X_k \mid \rm{PA}_k}.\] 
For example, in \cref{sec:examp} we considered $P^{\langle 1, 0 \rangle}_{Y}$, where we shifted $P_X$ to be as in sample 1 but kept $P_{Y \mid X}$ to be as in sample 0. Of all the possible $P^{\bm{c}}_Y$ for $\bm{c} \in \{0,1\}^{K+1}$, only two are observed directly in the data: $P^{(0)}_{Y} := P^{\langle 0, \ldots, 0, 0\rangle}_{Y}$ and $P^{(1)}_{Y} := P^{\langle 1, \ldots, 1, 1\rangle}_{Y}$; the rest are counterfactual distributions. 

We will quantify differences in the distribution of the outcome through the change in some functional (summary measure) of the form $\theta(P_Y) = \int h(y) \rm{d}P_{Y}(y)$ for some $h \colon \RR \to \RR$. Important examples of such functionals are the mean $h(y) = y$, the second moment $h(y) = y^2$ (which, combined with the mean, allows to obtain the variance), and the CDF at a point $u$: $h_u(y) = \ind\{y \leq u\}$ (which can be inverted to obtain quantiles and Wasserstein distances between two distributions). Our main results concern identification, estimation and inference on $\theta^{\bm{c}} : = \theta(P_Y^{\bm c})$ for a counterfactual distribution described by change vector $\bm{c} \in \{0,1\}^{K+1}$.

To state our main results, we adopt the following notation. We denote with a bar $\Bar{\bm{X}}_{k} := (X_1, \ldots, X_k)$ for any $k \leq K$. In a slight abuse of notation, we write $\Bar{\bm{X}}_{K+1} = (\bm{X}, Y)$ and define $g_{K+1}(\Bar{\bm{X}}_{K+1}) := h(Y)$ as a convention to make mathematical expressions more compact.

\begin{theorem}\label{thm:main}
Let $g_k(\Bar{\bm{X}}_k)$, $a_k(\Bar{\bm{X}}_k)$ be any candidate functions such that $\rm{E}_{(c_{k+1})}[g_k(\Bar{\bm{X}}_k)^2] < \infty$, $\rm{E}_{(c_{k+1})}[a_k(\Bar{\bm{X}}_k)^2] < \infty$ for $k = 1, \ldots, K$.
Consider the following estimating equation:
\begin{multline}
    \rm{E}_{(c_1)}[g_1(X_1)] + \sum_{k=1}^K \rm{E}_{(c_{k+1})}[a_k(\Bar{\bm{X}}_k) e_k(\Bar{\bm{X}}_{k+1})], \tag{MR} \label{eq:MR}
\end{multline}
where $e_k(\Bar{\bm{X}}_{k+1}) : = g_{k+1}(\Bar{\bm{X}}_{k+1}) - g_{k}(\Bar{\bm{X}}_k)$ for $k = 1, \ldots, K$.

For $k = 1, \ldots, K$ define:
\begin{align*}
 \gamma_k(\Bar{\bm{X}}_k) & := \rm{E}_{(c_{k+1})} [g_{k+1}(\Bar{\bm{X}}_{k+1}) \mid \Bar{\bm{X}}_k], \\
\alpha_k(\Bar{\bm{X}}_k) & := \prod_{j=1}^k \frac{\rm{d}P^{(c_j)}_{X_{j} \mid \Bar{\bm{X}}_{j-1}}}{\rm{d}P^{(c_{k+1})}_{X_{j} \mid \Bar{\bm{X}}_{j-1}}} (X_{j} \mid \Bar{\bm{X}}_{j-1}). 
\end{align*}
Under regularity conditions given in the appendix, \eqref{eq:MR} is equal to $\theta^{\bm c}$ if, for every $k = 1,\ldots, K$, either $g_k(\Bar{\bm{X}}_k) = \gamma_k(\Bar{\bm{X}}_k)$ or $a_k(\Bar{\bm{X}}_k) = \alpha_k(\Bar{\bm{X}}_k)$, but not necessarily both. 
\end{theorem}

The intuition for this result is similar to \cref{lem:exampleDR}. The parameter of interest can be estimated by \emph{regression}, but now we need to use sequentially nested regressions $\gamma_k(\Bar{\bm{X}}_k)$ (i.e. regressions where the outcome itself is a regression function). This fixes the desired distribution for each causal mechanism. Since we are estimating $K$ regression functions, we require $K$ debiasing terms, which appear additively. Each of these debiasing terms includes a ``weight'' with true value $\alpha_k(\Bar{\bm{X}}_k)$, which is a product of the RN derivatives for the distributions that are shifted with respect to $c_{k+1}$. Including these $K$ debiasing terms makes equation \eqref{eq:MR} \emph{multiply robust}: for each causal mechanism, only the corresponding regression function or the corresponding weights need to be correctly specified, but not necessarily both. We also note that one of our conditions for multiple robustness is that $g_k(\Bar{\bm{X}}_k) = \gamma_k(\Bar{\bm{X}}_k)$ where $\gamma_k(\Bar{\bm{X}}_k) := \rm{E}_{(c_{k+1})} [g_{k+1}(\Bar{\bm{X}}_{k+1}) \mid \Bar{\bm{X}}_k]$. This is weaker than setting $\gamma_k(\Bar{\bm{X}}_k) = \rm{E}_{(c_{k+1})} [\gamma_{k+1}(\Bar{\bm{X}}_{k+1}) \mid \Bar{\bm{X}}_k]$, because we do not require that $g_{k+1}$ is correctly specified, as long as $a_{k+1}(\Bar{\bm{X}}_{k+1}) = \alpha_{k+1}(\Bar{\bm{X}}_{k+1})$.

\begin{example}\label{ex:mediat}
Suppose $K=2$ and the causal Markov factorization \eqref{eq:factor} is $P_{Y \mid X_1, X_2} P_{X_2 \mid X_1} P_{X_1}$, as captured by the DAG in \cref{fig:dag_examp}. Suppose we are interested in estimating $\theta^{\bm{c}}$ for $\bm{c} = \langle 0, 1, 0 \rangle$. The multiply-robust estimating equation in \cref{thm:main} for this case is:
\begin{multline*}
    \rm{E}_{(0)}[g_1(X_1)] \\ + \rm{E}_{(1)}[a_1(X_1)\{g_2(X_1, X_2) - g_1(X_1)\}] \\ +  \rm{E}_{(0)}[a_2(X_1, X_2)\{h(Y) - g_2(X_1, X_2)\}],
\end{multline*}
where the true values of the unknown functions are:
\begin{align*}
    \gamma_2(X_1, X_2) & = \rm{E}_{(0)}[h(Y) \mid X_1, X_2], \\
    \gamma_1(X_1) & = \rm{E}_{(1)}[g_2(X_1, X_2) \mid X_1], \\
    \alpha_2(X_1, X_2) & = \rm{d}P^{(1)}_{X_2 \mid X_1}/ \rm{d}P^{(0)}_{X_2 \mid X_1} (X_2 \mid X_1), \\ 
    \alpha_1(X_1) & = \rm{d}P^{(0)}_{X_1}/ \rm{d}P^{(1)}_{X_1} (X_1). \qedhere
\end{align*}
\end{example}

\begin{figure}[!h]
    \centering
    \input{figures/dag_examp}
    \caption{DAG for \cref{ex:mediat}.}
    \label{fig:dag_examp}
\end{figure}

\begin{remark}[Identification with Covariates] As discussed in \cref{sec:structural}, under the assumption that $T$ is a randomly assigned treatment, the causal change attribution problem has an interpretation in terms of mediation \citet[\S4.5]{pearl2009causality}. Sometimes, random assignment of $T$ may be implausible, but the researcher may have access to a set of pre-treatment covariates $\bm W$ such that an unconfoundedness assumption holds conditional on $\bm W$. It is still possible to get multiple robustness in that case, with two adjustments: (i) the regression and weights need to control for $\bm W$, and (ii) an additional debiasing term is required, to account for the adjustment on covariates. This is explained in \cref{sec:covariates}.
\end{remark}

\begin{remark}[Runtime and Speedup]
The general result of \cref{thm:main} implies that, when there are $K$ explanatory variables, we can identify $\theta^{\bm c}$ for any $\bm{c}$ with a multiply-robust estimating equation that depends on $2K$ unknown functions ($K$ regression functions and $K$ weights) that need to be estimated. For certain change vectors $\bm{c}$, however, it is possible to simplify the identifying equation to have fewer unknown functions while maintaining multiple robustness. In \cref{sec:simpl}, we discuss two settings where simplifications may occur. The first one is when $c_k = c_{k+1}$ for some $k$ (i.e., when two consecutive regressions are computed with respect to the same probability distribution). The second one is when $X_k$ and $X_{k+1}$ are (conditionally) independent. In particular, when all the explanatory variables are independent, for any given $\bm{c}$ we only need to estimate one regression and one RN derivative, regardless of $K$.
\end{remark}

\subsection{Estimation}\label{sec:est}
In this section, we describe how to implement the multiply-robust estimating equation \eqref{eq:MR} in practice. 

\paragraph{Step 1. Estimate the Regressions} The nested regression functions $\gamma_k(\Bar{\bm{X}}_k)$ can be estimated by any parametric or non-parametric ML regression algorithm, including LASSO, ridge, random forests, neural networks, boosting, or any ensemble method combining multiple of these. We work recursively, starting with an estimator $\hat \gamma_K(\Bar{\bm{X}}_K)$ for the regression of $h(Y)$ on $\Bar{\bm{X}}_K := \bm{X}$, using the the data from sample $c_{K+1} \in \{0,1\}$ (i.e., the distribution we want to fix for $P_{Y \mid \bm X}$). Next, we obtain $\hat \gamma_{K-1}(\Bar{\bm{X}}_{K-1})$ by regressing $\hat \gamma_K(\Bar{\bm{X}}_K)$ on $\Bar{\bm{X}}_{K-1}$, using the the data from sample $c_{K} \in \{0,1\}$ (i.e., the distribution we want to fix for $P_{X_K \mid \rm{PA}_K}$). We proceed this way up until we obtain $\hat \gamma_1(X_1)$.

\paragraph{Step 2. Estimate the Weights} We propose two different approaches to estimate the RN derivatives that appear in the weights $\alpha_k(\Bar{\bm{X}}_k)$: \emph{classification} and \emph{automatic estimation}. Both methods target the weights directly, rather than estimating a density or probability mass function in each sample and then taking the ratio. For conciseness, we only describe estimation of $\mu(\Bar{\bm{X}}_j) := \rm{d}P^{(1)}_{\Bar{\bm{X}}_j}/\rm{d}P^{(0)}_{\Bar{\bm{X}}_j}(\Bar{\bm{X}}_j)$. The reciprocal RN derivative 
can be estimated analogously by exchanging the indices 0 and 1. Conditional RN derivatives for $X_j \mid \Bar{\bm{X}}_{j-1}$ can be obtained by dividing the RN derivative for $\Bar{\bm{X}}_j$ by the RN derivative for $\Bar{\bm{X}}_{j-1}$. 

\paragraph{2a. Classification} By Bayes' rule, we can express:
\begin{equation}
 \mu(\Bar{\bm{X}}_j) = \frac{\beta(\Bar{\bm{X}}_j)}{1 - \beta(\Bar{\bm{X}}_j)} \frac{1-p}{p}, \tag{BAY} \label{eq:bayes} 
\end{equation}
where $\beta(\Bar{\bm{X}}_j) := \Pr(T = 1 \mid \Bar{\bm{X}}_j)$ and $p := \Pr(T = 1)$ (the posterior and prior probabilities of $\Bar{\bm{X}}_j$ coming from sample 1, respectively). For a given $p$, this defines a one-to-one mapping between $\mu(\Bar{\bm{X}}_j)$ and $\beta(\Bar{\bm{X}}_j)$.

This suggests the following methodology to estimate $\mu(\Bar{\bm{X}}_j)$. First, we train a classification ML algorithm (logistic regression with LASSO or ridge penalties, neural networks, random forests, etc.) to predict $T$ based on the concatenated data for $\Bar{\bm{X}}_j$. Second, we replace $\beta(\Bar{\bm{X}}_j)$ in Bayes' rule with the predicted posterior probabilities, and we replace $(1-p)/p$ by its empirical analog, $n_0/n_1$, where $n_t$ is the number of observations in the $T = t$ sample. The posterior probabilities can also be calibrated by cross-validation, as discussed, e.g., in \citet{niculescu2005predicting}.

Although this method of estimating RN derivatives is not new (c.f., for instance, \citealp[pt. II, ch. 4]{sugiyama2012density}, or \citealp{arbour2021permutation}), the application to building a multiply-robust moment function is novel.

\paragraph{2b. Automatic Estimation} We specialize general results derived by \citet{chernozhukov2021automatic} to the causal change attribution problem. In particular, we can characterize the RN derivative $\mu(\Bar{\bm{X}}_j)$ as the solution of:
\[\mu(\Bar{\bm{X}}_j) = \arg\min_m  \rm{E}_{(0)}[m(\Bar{\bm{X}}_j)^2] - 2\rm{E}_{(1)}[m(\Bar{\bm{X}}_j)]. \]
The literature refers to this method as ``automatic'' estimation of the weights because the loss function above depends only in $m(\Bar{\bm{X}}_j)$, without using the explicit form of $\mu(\Bar{\bm{X}}_j)$ as an RN derivative. Thus, we could minimize a sample version of the criterion above over some class of functions (e.g. linear with LASSO or ridge penalties, neural networks, random forests, etc.).

\paragraph{Step 3. Estimate the Target Parameter} Finally, we build an estimator of the counterfactual parameter $\theta^{\bm{c}}$ by replacing $g(\Bar{\bm{X}}_k)$ with $\hat\gamma_k(\Bar{\bm{X}}_k)$ and $a(\Bar{\bm{X}}_k)$ with $\hat\alpha_k(\Bar{\bm{X}}_k)$ in a sample analog of equation \eqref{eq:MR}:
\begin{multline}
  \hat\theta^{\bm c} := \widehat{\rm{E}}_{(c_1)} [\hat{\gamma}_1(X_{1})] \\  + \sum_{k=1}^{K} \widehat{\rm{E}}_{(c_{k+1})} [\hat{\alpha}_k(\Bar{\bm{X}}_{k}) \hat{e}_k(\Bar{\bm{X}}_{k+1})], \tag{EST}\label{eq:est}
\end{multline}
where $\hat{e}_k(\Bar{\bm{X}}_{i,k+1}) : = \hat\gamma_{k+1}(\Bar{\bm{X}}_{k+1}) - \hat\gamma_{k}(\Bar{\bm{X}}_k)$ for $k = 1, \ldots, K$, and we use $\widehat{\rm{E}}_{(t)}[f(W)] = n_t^{-1} \sum_{i : T_i = t} f(W_i)$ to denote the sample average on sample $t \in \{0, 1\}$.

\begin{remark}[Sample-splitting]
To avoid overfitting bias, our recommendation is that the data used to learn the unknown functions $\gamma_k(\Bar{\bm{X}}_k)$, $\alpha_k(\Bar{\bm{X}}_k)$ are not re-used to estimate the main parameter $\theta^{\bm{c}}$. When the dataset is large, different random subsamples can be used. In medium-sized datasets, a good alternative is to use cross-fitting (see, for example, \citealp{newey2018cross}).
\end{remark}

\subsection{Large-Sample Inference}\label{sec:inference}
In this section, we show consistency and asymptotic normality of our estimator $\hat\theta^{\bm c}$, and explain how to perform large-sample inference on the true parameter $\theta^{\bm c}$. First, we notice that the estimator in \eqref{eq:est} is the sum of two averages over different samples:
\[\hat\theta^{\bm c} = \widehat{\rm{E}}_{(0)}[\hat\psi_{(0)}(\bar{\bm{X}}_{K+1})] + \widehat{\rm{E}}_{(1)}[\hat\psi_{(1)}(\bar{\bm{X}}_{K+1})],\]
where $\hat\psi_{(t)}(\bar{\bm{X}}_{K+1})$ contains the terms that depend on
the data from sample $t \in \{0,1\}$ (the exact expression is given in the proof of \cref{thm:asymp}). This characterization will be useful in the results below in terms of writing the asymptotic variance of $\hat\theta^{\bm c}$. We define an ``oracle'' version of the same object, $\psi_{(t)}(\bar{\bm{X}}_{K+1})$, by replacing $\hat\gamma_k(\bar{\bm{X}}_k), \hat\alpha_k(\bar{\bm{X}}_k)$ with their respective true values $\gamma_k(\bar{\bm{X}}_k), \hat\alpha_k(\bar{\bm{X}}_k)$ for $k = 1, \ldots, K$.

The following theorem gives an asymptotic normality result for  $\hat\theta^{\bm c}$, which allow us to perform valid inference on the true parameter $\theta^{\bm c}$ in large samples (e.g., provide standard errors or confidence intervals):

\begin{theorem}\label{thm:asymp}
Under regularity conditions given in the appendix, $\hat\theta^{\bm c}$ is consistent and asymptotically normal:
\begin{gather*}
    \hat\theta^{\bm c}\overset{p}{\longrightarrow} \theta^{\bm c} \quad \text{and} \quad \sqrt{n} (\hat\theta^{\bm c}- \theta^{\bm c}) \overset{d}{\longrightarrow} N(0,V),
\end{gather*} where $n := n_0 + n_1$ and 
\[V := \frac{1}{1-p}\rm{Var}_{(0)}[\psi_{(0)}(\bar{\bm{X}}_{K+1})] + \frac{1}{p}\rm{Var}_{(1)}[\psi_{(1)}(\bar{\bm{X}}_{K+1})].\]
Moreover, $V$ can be estimated consistently by:
\[\hat V := \frac{n}{n_0}\widehat{\rm{Var}}_{(0)}[\hat\psi_{(0)}(\bar{\bm{X}}_{K+1})] +  \frac{n}{n_1}\widehat{\rm{Var}}_{(1)}[\hat\psi_{(1)}(\bar{\bm{X}}_{K+1})]\]
and
$\rm{Pr}(\theta^{\bm c} \in [\hat\theta^{\bm c}\mp z_{1-a/2} \times (\hat{V}/n)^{1/2}]) \to 1 - a$, where $z_{1-a/2}$ be the $(1-a/2)$-th quantile of a standard Gaussian random variable.
\end{theorem}


\begin{remark}[On the Rate Conditions]\label{rem:rates}
\cref{ass:rates}, given in the appendix, imposes certain requirements on the root mean-square (RMSE) convergence rates of $\hat\gamma_k$ and $\hat\alpha_k$, which depend on the choice of algorithm and the properties of the true regression and weighting functions (smoothness, sparsity, etc.). The main condition is that the product of the RMSE for the regression and for the weights has to vanish faster than $n^{-1/2}$. This restriction guarantees that the bias of the main estimator $\hat\theta^{\bm c}$ is small enough in large samples to obtain valid inference. The product condition captures an important trade-off: in situations where the regression functions can be estimated at a fast rate, the method will work even if the weights converge very slowly, and vice versa, which is a key advantage of using a multiply-robust estimating equation.
\end{remark}

\subsection{Quantifying the Contribution of Each Causal Mechanism}\label{sec:combin}
We now discuss two ways to use the $\theta^{\bm c}$ for the problem of causal attribution: \emph{Shapley values} and \emph{along a causal path}.

\paragraph{Shapley Values} 
Shapley values compute the effect of shifting a causal mechanism averaging over each possible combination of which other causal mechanisms to change. Formally, let $\bm e_k$ be a $(K+1)$-vector with a 1 in the $k$-th entry and 0 everywhere else. The Shapley value associated with $P_{X_k \mid \rm{PA}_k}$ is:
\[\rm{SHAP}_k = \sum_{\bm{c} : c_k = 0} \frac{1}{(K+1) {\binom{K}{\sum_j c_j}}} (\theta^{\bm{c} + \bm{e}_k} - \theta^{\bm{c}}).\]

\paragraph{Along a Causal Path} Define, for each $k = 1, \ldots, K+1$, a vector $\bm{b}_k \in \{0,1\}^{K+1}$ with entries $1, \ldots, k$ equal to 1, and the rest equal to 0. In this method, we define the contribution of the $k$-th causal mechanism to be:
\[\rm{PATH}_k = \theta^{\bm b_k} - \theta^{\bm b_{k-1}}.\]
In other words, we define the contribution of the $k$-th causal mechanism as the effect of shifting from $P_{X_k \mid \rm{PA}_k}$ fixing its causal antecedents to be as in sample 1, but its causal descendants to be as in sample 0. 

\paragraph{How to choose between these two approaches?} As discussed in the literature \citep{budhathoki2021did}, the effect of changing the one causal mechanism may depend on which other mechanisms have already been changed; Shapley values take that into account, whereas the method along a causal path does not. Whether this is a relevant concern depends on the application: in some cases, the impact of a given explanatory variable may be approximately the same regardless of the distribution of other explanatory variables. On the other hand, causal attribution along a causal path requires estimating $\theta^{\bm c}$ for only $K$ change vectors $\bm{c}$ (and, by the simplification in \cref{rem:iter}, each requires training only one regressor and one classifier), whereas computing Shapley values requires estimating $\theta^{\bm c}$ for $2^{K+1}$ change vectors $\bm{c}$ (each of which could require training up to $2K$ machine learners). Some computationally efficient approximations to $\rm{SHAP}_k$ exist (e.g, \citealp{kolpaczki2023approximating}).

Let $\widehat{\rm{SHAP}}_k$ and $\widehat{\rm{PATH}}_k$ be estimators of $\rm{SHAP}_k$ and $\rm{PATH}_k$, respectively, that replace the true $\theta^{\bm c}$ with the multiply-robust estimator $\hat{\theta}^{\bm c}$ in \eqref{eq:est}. The following corollary gives a useful result for large-sample inference on the causal attribution parameters:
\begin{corollary}
Under the conditions of \cref{thm:asymp}, $\widehat{\rm{SHAP}}_k$ and $\widehat{\rm{PATH}}_k$ are consistent and asymptotically normal, with a consistently estimable asymptotic variance.
\end{corollary}

This is a consequence of $\widehat{\rm{SHAP}}_k$ and $\widehat{\rm{PATH}}_k$ being finite linear combinations of $\hat{\theta}^{\bm c}$. The asymptotic variance could be computed explicitly, although this might be cumbersome, especially for Shapley values, since it requires the $2^{K+1}\times 2^{K+1}$ covariance matrix of $\hat{\theta}^{\bm c}$ for $\bm c \in \{0,1\}^{K+1}$. As an alternative, the multiplier bootstrap of \citet{belloni2017program} can be used. We describe the procedure in \cref{sec:boot}. An advantage of the multiplier approach is that each bootstrap replication does not require re-training the ML estimators of the regression function and weights.

\section{Empirical Results}\label{sec:empirical}
In this section, we discuss two sets of Monte-Carlo simulation results and a real-data application of our method to understanding the gender wage gap. The code for the simulations and the empirical application is available at \url{https://github.com/victor5as/mr_causal_attribution}.

\subsection{Monte-Carlo Simulations}\label{sec:mc}
\paragraph{Design 1}
We now present simulation evidence on the performance of our method. Our first simulation design is based on the causal model of \cref{ex:mediat}, with DAG $X_1 \rightarrow X_2$, $X_1 \rightarrow Y$ and $X_2 \rightarrow Y$. We give details about the data-generating process in \cref{sec:appmc}.

\cref{tab:mcres} presents the results of our simulation (over 1,000 draws) for different choices of learners for the regression and the weights. \ref{tab:mca} shows the Mean Absolute Error (MAE) for a scenario where both the regression and the weights are correctly specified parametric learners: OLS with quadratic terms for the regressions and logistic classifier with quadratic terms to build the weights.\footnote{This is the correct specification for the weights, since the log-likelihood ratio between two Gaussian distributions with possibly different variances is a quadratic polynomial.} Comparing our multiply-robust (MR) estimator to using regression or re-weighting only, we can see that its performance is statistically indistinguishable from the best of the two methods (regression). 

Tables \ref{tab:mcb} and \ref{tab:mcc} show what happens when either the weights or the regressions are misspecified, while the specification of the other component of the model is still correct. As a particular form of misspecification, we omit the quadratic terms from either the OLS regression or the logistic classifier. In either case, we observe that MR performs as well as the correctly-specified method, while the misspecified method exhibits much higher errors. This illustrates the multiple robustness property of our estimator. When both the weights and the regression are misspecified parametric models, as in \ref{tab:mcd}, our theoretical guarantees no longer apply. However, we still show these results for completeness. Although it seems to be the case that the error in Shapley values achieved by the MR methods is about as low as the best of the two other methods, it is significantly larger than the correctly-specified benchmark of \cref{tab:mca}.

Next, we turn our attention to flexible, non-parametric learners for the unknown regression functions and weights, to mimic a more realistic situation in which we do not want to make strong parametric assumptions about these. First, we consider neural networks (\cref{tab:mce}). We use \texttt{MLPRegressor} and \texttt{MLP\-Classifier} from the Python library \texttt{sk\-learn}. We use the default settings for all hyperparameters, except that we allow for early stopping and use 3 hidden layers of 100 neurons each. Second, we consider gradient boosting (\cref{tab:mcf}), using \texttt{Gradient\-Boosting\-Regressor} and \texttt{Gradient\-Boosting\-Classifier} from \texttt{sk\-learn} with the default settings. In both cases, we calibrate the predicted probabilities after classification with \texttt{CalibratedClassifierCV} also from \texttt{sk\-learn}. The results clearly illustrate \cref{rem:rates}: the bias of the MR non-parametric estimator will, in general, vanish faster than the convergence rates of regression-only or re-weighting-only methods. As such, we see that in general MR has equal or lower error than the best of the other two methods. Moreover, the MAE is often statistically indistinguishable from that of the correctly-specified parametric benchmark \ref{tab:mca}.

\paragraph{Design 2} In a second set of experiments, we consider the effect of increasing hte number of explanatory variables $K$. We consider a line DAG, $X_1 \rightarrow X_2 \rightarrow \cdots \rightarrow X_K \rightarrow Y$. We give more details about the data-generating process in \cref{sec:appmc}. We use LASSO to learn the regression functions, and Logistic Regression with $\ell_1$ penalty (Logit-LASSO) for the weights. In both cases, the penalty is selected by cross-validation.

\cref{tab:mc2} shows the average (across 500 simulations) of the worst-case (across parameters) absolute error, that is, the average of $\max_{k = 1, \ldots, K+1} |\rm{PATH}_k - \widehat{\rm{PATH}}_k|$. The pattern is almost identical if we look at the Mean Absolute Error instead. The MR estimator performs significantly better than the other two methods, even though the regression and the weights are correctly specified (i.e., they are truly linear/logistic and sparse).

\subsection{Real-World Application: Gender Wage Gap}\label{sec:appli}
Following \citet{budhathoki2021did}, we ask how much of the gender wage gap can be attributed to differences in education or occupation. We use data from the Current Population Survey (CPS) 2015. After applying the same sample restrictions as \citet{chernozhukov2018sorted}, the resulting sample contains 18,137 male and 14,382 female employed individuals. On this data, the (unconditional) gender wage gap in this sample is of \$7.91/hour (standard error 0.01). In other words, female workers earn 24\% less on average than male workers; the difference is statistically significant.

\begin{figure}[!t]
    \centering
    \input{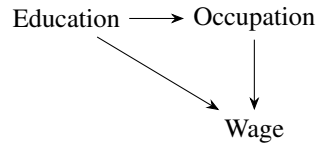}
    \caption{DAG for the gender wage gap application.}
    \label{fig:dag_gendergap}
\end{figure}

We assume the same causal graphical model as \citet{budhathoki2021did}, which is captured by the DAG in \cref{fig:dag_gendergap}. We randomly split the data into a training set used to estimate the regression function and weights (40\%), a calibration set to calibrate the predicted probabilities from our classification algorithm (10\%), and an evaluation set used to obtain the main estimates of the $\hat\theta^{\bm c}$ parameters (50\%). To estimate the regression and the weights, we use \texttt{Hist\-Gradient\-Boosting\-Regressor} and \texttt{Hist\-Gradient\-Boosting\-Classifier} from \texttt{sklearn} in Python. We calibrate the probabilities by isotonic regression.

\begin{figure*}[!tbhp]
    \centering
    \includegraphics[width=0.9\textwidth]{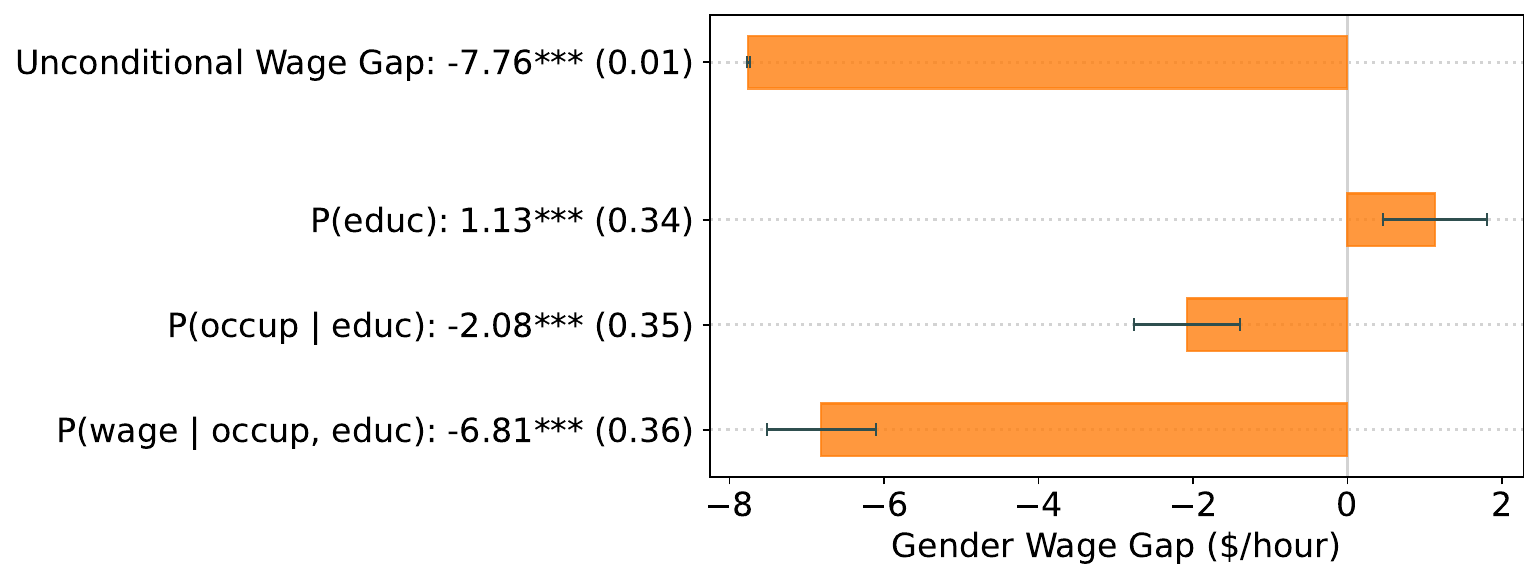}
    \caption{Shapley Values in our gender wage gap application. We plot the point estimates and 95\% confidence intervals. We also report the point estimates (bootstrapped standard errors in brackets). We denote statistical significance at the 5\% level by $^{**}$, and at the 1\% level by $^{***}$.}
    \label{fig:appli_shap}
\end{figure*}

We plot our estimated Shapley values are in \cref{fig:appli_shap}. The contribution of education is estimated to be positive and significant. In contrast, the distribution of occupation given education has a significant negative effect, slightly larger in magnitude than the contribution of education, but of opposite sign. Most of the gender wage gap is therefore explained by differences in the distribution of wages given education and occupation. This is consistent with previous findings in the literature. For example, \citet{chernozhukov2018distribution}, documents the same fact for the UK. 

One could think of $P(\text{wage} \mid \text{occup, educ})$ as ``residual'' variation, i.e., the remaining gender gap which is not explained by education or occupation. It may capture the effect of additional variables that are not included in our analysis. For example, there may be gender differences in labor market experience due to women temporarily exiting the labor force to take care of young children. Unfortunately, experience is not available in census data, and so we cannot quantify its impact, but excluding it from the analysis does not bias our results, because it is plausibly not a causal antecedent of education and occupational choice. It may also be due to other factors, including gender discrimination in promotion or gender differences in wage bargaining.

To gain more insight into the results of our attribution exercise, it is natural to ask two questions: (1) how different is the distribution of education and occupation between the two groups, and (2) how do these differences translate into differences in average wage. We present some summary statistics related to these questions in the appendix. \cref{fig:educ} compares the distribution of educational attainment for males and females. It appears that females tend to be slightly more educated, with higher rates of college and advanced degree graduation. This could explain the positive Shapley value associated with education; for comparison, college graduates earn \$12.60/hour more than high school graduates on average. \cref{fig:occup} shows the distribution of occupation conditional on having a college degree. Females are more predominant in administrative, education or healthcare professions, whereas males are more likely to work in management and sales. For comparison, managers earn \$16.66/hour more than educators and \$6.29/hour more than healthcare practitioners on average. At the same time, there are wide differences across genders that cannot be explained by education and occupation alone: female college graduate managers earn \$13.77/hour less on average than their male counterparts, which is consistent with the finding that $P(\text{wage} \mid \text{occup, educ})$ has the largest Shapley value in our causal change attribution exercise.

\section{Conclusions}\label{sec:conclusions}
In this paper, we develop a new estimator for causal change attribution measures, which combines regression and re-weighting methods in a multiply-robust estimating equation. We provide the first results on consistency and asymptotic normality of ML-based estimators in this setting, and discuss how to perform inference. Moreover, our method can be used to estimate Shapley values, which will inherit the consistency and large-sample distribution properties. 

Finally, we suggest one direction for future research. The causal interpretation of the change attribution parameters relies on an assumption of no unobserved confounding. The sensitivity bounds in \citet{Chernozhukov2022long} could be adapted as a way to test the robustness of a causal change attribution study to unobserved confounding.

\section*{Acknowledgements}
The authors would like to thank Victor Chernozhukov, Dominik Janzing, Eric Tchetgen-Tchetgen, and seminar audiences at Amazon and MIT for providing helpful feedback. We also would like to thank Patrick Blöbaum for help with code integration in \texttt{DoWhy}.


\section*{Impact Statement}
This paper presents work whose goal is to advance the fields of Causal Inference and Machine Learning. There are many potential societal consequences of our work, none which we feel must be specifically highlighted here.

The ``causal'' interpretation of our change attribution measures is grounded on certain assumptions about the Directed Acyclic Graph (DAG) that codifies causal dependence among the variables in the data. We are precise about these assumptions (\cref{sec:setting}), and discuss some ways to relax them in future extensions (\cref{sec:conclusions}). Our theoretical results also rely on regularity conditions, which we state formally in the Appendix and explain in more intuitive terms in the main body of the paper. Practitioners need to be mindful that, if these assumptions do not hold, our proposed algorithm is not guaranteed to lead to accurate results.

\bibliography{references}

\begin{thebibliography}{45}
\providecommand{\natexlab}[1]{#1}
\providecommand{\url}[1]{\texttt{#1}}
\expandafter\ifx\csname urlstyle\endcsname\relax
  \providecommand{\doi}[1]{doi: #1}\else
  \providecommand{\doi}{doi: \begingroup \urlstyle{rm}\Url}\fi

\bibitem[Arbour et~al.(2021)Arbour, Dimmery, and Sondhi]{arbour2021permutation}
Arbour, D., Dimmery, D., and Sondhi, A.
\newblock Permutation weighting.
\newblock In \emph{International Conference on Machine Learning}, pp.\
  331--341. PMLR, 2021.

\bibitem[Belloni et~al.(2017)Belloni, Chernozhukov, Fern{\'a}ndez-Val, and
  Hansen]{belloni2017program}
Belloni, A., Chernozhukov, V., Fern{\'a}ndez-Val, I., and Hansen, C.
\newblock Program evaluation and causal inference with high-dimensional data.
\newblock \emph{Econometrica}, 85\penalty0 (1):\penalty0 233--298, 2017.

\bibitem[Berman(2018)]{berman2018beyond}
Berman, R.
\newblock Beyond the last touch: Attribution in online advertising.
\newblock \emph{Marketing Science}, 37\penalty0 (5):\penalty0 771--792, 2018.

\bibitem[Bickel et~al.(2009)Bickel, Br{\"u}ckner, and
  Scheffer]{bickel2009discriminative}
Bickel, S., Br{\"u}ckner, M., and Scheffer, T.
\newblock Discriminative learning under covariate shift.
\newblock \emph{Journal of Machine Learning Research}, 10\penalty0 (9), 2009.

\bibitem[Billingsley(1995)]{billingsley1995probability}
Billingsley, P.
\newblock \emph{Probability and Measure}.
\newblock John Wiley \& Sons, third edition, 1995.

\bibitem[Blinder(1973)]{blinder1973wage}
Blinder, A.~S.
\newblock Wage discrimination: reduced form and structural estimates.
\newblock \emph{Journal of Human resources}, pp.\  436--455, 1973.

\bibitem[Bl{\"o}baum et~al.(2022)Bl{\"o}baum, G{\"o}tz, Budhathoki, Mastakouri,
  and Janzing]{dowhy_gcm}
Bl{\"o}baum, P., G{\"o}tz, P., Budhathoki, K., Mastakouri, A.~A., and Janzing,
  D.
\newblock Dowhy-gcm: An extension of dowhy for causal inference in graphical
  causal models.
\newblock \emph{arXiv preprint arXiv:2206.06821}, 2022.

\bibitem[Budhathoki et~al.(2021)Budhathoki, Janzing, Bloebaum, and
  Ng]{budhathoki2021did}
Budhathoki, K., Janzing, D., Bloebaum, P., and Ng, H.
\newblock Why did the distribution change?
\newblock In \emph{AISTATS}, 2021.

\bibitem[Budhathoki et~al.(2022)Budhathoki, Minorics, Bl{\"o}baum, and
  Janzing]{budhathoki2022causal}
Budhathoki, K., Minorics, L., Bl{\"o}baum, P., and Janzing, D.
\newblock Causal structure based root cause analysis of outliers.
\newblock In \emph{International Conference on Machine Learning}, pp.\
  2357--2369. PMLR, 2022.

\bibitem[Chen et~al.(2023)Chen, Covert, Lundberg, and Lee]{chen2023algorithms}
Chen, H., Covert, I.~C., Lundberg, S.~M., and Lee, S.-I.
\newblock Algorithms to estimate shapley value feature attributions.
\newblock \emph{Nature Machine Intelligence}, pp.\  1--12, 2023.

\bibitem[Chernozhukov et~al.(2018{\natexlab{a}})Chernozhukov, Chetverikov,
  Demirer, Duflo, Hansen, Newey, and Robins]{chernozhukov2018double}
Chernozhukov, V., Chetverikov, D., Demirer, M., Duflo, E., Hansen, C., Newey,
  W., and Robins, J.
\newblock Double/debiased machine learning for treatment and structural
  parameters.
\newblock \emph{The Econometrics Journal}, 21\penalty0 (1):\penalty0 C1--C68,
  2018{\natexlab{a}}.

\bibitem[Chernozhukov et~al.(2018{\natexlab{b}})Chernozhukov, Fern\'andez-Val,
  and Luo]{chernozhukov2018distribution}
Chernozhukov, V., Fern\'andez-Val, I., and Luo, S.
\newblock Distribution regression with sample selection, with an application to
  wage decompositions in the {UK}.
\newblock Technical Report 1811.11603, arXiv.org, 2018{\natexlab{b}}.

\bibitem[Chernozhukov et~al.(2018{\natexlab{c}})Chernozhukov,
  Fern{\'a}ndez-Val, and Luo]{chernozhukov2018sorted}
Chernozhukov, V., Fern{\'a}ndez-Val, I., and Luo, Y.
\newblock The sorted effects method: Discovering heterogeneous effects beyond
  their averages.
\newblock \emph{Econometrica}, 86\penalty0 (6):\penalty0 1911--1938,
  2018{\natexlab{c}}.

\bibitem[Chernozhukov et~al.(2021)Chernozhukov, Newey, Quintas-Mart\'inez, and
  Syrgkanis]{chernozhukov2021automatic}
Chernozhukov, V., Newey, W.~K., Quintas-Mart\'inez, V., and Syrgkanis, V.
\newblock Automatic debiased machine learning via {R}iesz regression.
\newblock \emph{arXiv preprint arXiv:2104.14737}, 2021.

\bibitem[Chernozhukov et~al.(2022)Chernozhukov, Cinelli, Newey, Sharma, and
  Syrgkanis]{Chernozhukov2022long}
Chernozhukov, V., Cinelli, C., Newey, W., Sharma, A., and Syrgkanis, V.
\newblock {Long Story Short: Omitted Variable Bias in Causal Machine Learning}.
\newblock Technical report, National Bureau of Economic Research, Cambridge,
  MA, Jul 2022.

\bibitem[Chernozhukov et~al.(2023)Chernozhukov, Newey, Singh, and
  Syrgkanis]{chernozhukov2023automatic}
Chernozhukov, V., Newey, W., Singh, R., and Syrgkanis, V.
\newblock Automatic debiased machine learning for dynamic treatment effects and
  general nested functionals, 2023.

\bibitem[Dalessandro et~al.(2012)Dalessandro, Perlich, Stitelman, and
  Provost]{dalessandro2012causally}
Dalessandro, B., Perlich, C., Stitelman, O., and Provost, F.
\newblock Causally motivated attribution for online advertising.
\newblock In \emph{Proceedings of the sixth international workshop on data
  mining for online advertising and internet economy}, pp.\  1--9, 2012.

\bibitem[Daniel et~al.(2015)Daniel, De~Stavola, Cousens, and
  Vansteelandt]{daniel2015causal}
Daniel, R.~M., De~Stavola, B.~L., Cousens, S.~N., and Vansteelandt, S.
\newblock Causal mediation analysis with multiple mediators.
\newblock \emph{Biometrics}, 71\penalty0 (1):\penalty0 1--14, 2015.

\bibitem[Dawid et~al.(2014)Dawid, Faigman, and Fienberg]{dawid2014fitting}
Dawid, A.~P., Faigman, D.~L., and Fienberg, S.~E.
\newblock Fitting science into legal contexts: Assessing effects of causes or
  causes of effects?
\newblock \emph{Sociological Methods \& Research}, 43\penalty0 (3):\penalty0
  359--390, 2014.

\bibitem[Efron(2020)]{efron2020prediction}
Efron, B.
\newblock Prediction, estimation, and attribution.
\newblock \emph{International Statistical Review}, 88:\penalty0 S28--S59, 2020.

\bibitem[Imbens \& Rubin(2015)Imbens and Rubin]{imbens2015causal}
Imbens, G.~W. and Rubin, D.~B.
\newblock \emph{Causal inference in statistics, social, and biomedical
  sciences}.
\newblock Cambridge University Press, 2015.

\bibitem[Janzing et~al.(2024)Janzing, Bl{\"o}baum, Minorics, Faller, and
  Mastakouri]{janzing2024quantifying}
Janzing, D., Bl{\"o}baum, P., Minorics, L., Faller, P., and Mastakouri, A.
\newblock Quantifying intrinsic causal contributions via structure preserving
  interventions.
\newblock In \emph{International Conference on Artificial Intelligence and
  Statistics}. PMLR, 2024.

\bibitem[Ji et~al.(2016)Ji, Wang, and Zhang]{ji2016probabilistic}
Ji, W., Wang, X., and Zhang, D.
\newblock A probabilistic multi-touch attribution model for online advertising.
\newblock In \emph{Proceedings of the 25th acm international on conference on
  information and knowledge management}, pp.\  1373--1382, 2016.

\bibitem[Jung et~al.(2022)Jung, Kasiviswanathan, Tian, Janzing, Bl{\"o}baum,
  and Bareinboim]{jung2022measuring}
Jung, Y., Kasiviswanathan, S., Tian, J., Janzing, D., Bl{\"o}baum, P., and
  Bareinboim, E.
\newblock On measuring causal contributions via $do$-interventions.
\newblock In \emph{International Conference on Machine Learning}, pp.\
  10476--10501. PMLR, 2022.

\bibitem[Kolpaczki et~al.(2023)Kolpaczki, Bengs, Muschalik, and
  Hüllermeier]{kolpaczki2023approximating}
Kolpaczki, P., Bengs, V., Muschalik, M., and Hüllermeier, E.
\newblock Approximating the shapley value without marginal contributions.
\newblock Technical Report 2302.00736, arXiv.org, 2023.

\bibitem[Kulinski \& Inouye(2023)Kulinski and Inouye]{kulinski2023towards}
Kulinski, S. and Inouye, D.~I.
\newblock Towards explaining distribution shifts.
\newblock In \emph{International Conference on Machine Learning}, pp.\
  17931--17952. PMLR, 2023.

\bibitem[Liu et~al.(2023)Liu, Wang, Cui, and Namkoong]{liu2023need}
Liu, J., Wang, T., Cui, P., and Namkoong, H.
\newblock On the need for a language describing distribution shifts:
  Illustrations on tabular datasets.
\newblock \emph{arXiv preprint arXiv:2307.05284}, 2023.

\bibitem[Lu et~al.(2023)Lu, Geng, Li, Zhu, and Jia]{lu2023evaluating}
Lu, Z., Geng, Z., Li, W., Zhu, S., and Jia, J.
\newblock Evaluating causes of effects by posterior effects of causes.
\newblock \emph{Biometrika}, 110\penalty0 (2):\penalty0 449--465, 2023.

\bibitem[Mougan et~al.(2023)Mougan, Broelemann, Masip, Kasneci, Thiropanis, and
  Staab]{mougan2023explanation}
Mougan, C., Broelemann, K., Masip, D., Kasneci, G., Thiropanis, T., and Staab,
  S.
\newblock Explanation shift: Investigating interactions between models and
  shifting data distributions.
\newblock \emph{arXiv preprint arXiv:2303.08081}, 2023.

\bibitem[Newey \& Robins(2018)Newey and Robins]{newey2018cross}
Newey, W.~K. and Robins, J.~R.
\newblock Cross-fitting and fast remainder rates for semiparametric estimation.
\newblock \emph{arXiv preprint arXiv:1801.09138}, 2018.

\bibitem[Niculescu-Mizil \& Caruana(2005)Niculescu-Mizil and
  Caruana]{niculescu2005predicting}
Niculescu-Mizil, A. and Caruana, R.
\newblock Predicting good probabilities with supervised learning.
\newblock In \emph{Proceedings of the 22nd international conference on Machine
  learning}, pp.\  625--632, 2005.

\bibitem[Oaxaca(1973)]{oaxaca1973male}
Oaxaca, R.
\newblock Male-female wage differentials in urban labor markets.
\newblock \emph{International economic review}, pp.\  693--709, 1973.

\bibitem[Pearl(2009)]{pearl2009causality}
Pearl, J.
\newblock \emph{Causality}.
\newblock Cambridge University Press, 2009.

\bibitem[Peters et~al.(2017)Peters, Janzing, and
  Sch{\"o}lkopf]{peters2017elements}
Peters, J., Janzing, D., and Sch{\"o}lkopf, B.
\newblock \emph{Elements of causal inference: foundations and learning
  algorithms}.
\newblock The MIT Press, 2017.

\bibitem[Shao \& Li(2011)Shao and Li]{shao2011data}
Shao, X. and Li, L.
\newblock Data-driven multi-touch attribution models.
\newblock In \emph{Proceedings of the 17th ACM SIGKDD international conference
  on Knowledge discovery and data mining}, pp.\  258--264, 2011.

\bibitem[Shapley(1953)]{shapely1953value}
Shapley, L.
\newblock {A value for n-person games. Contributions to the theory of games},
  1953.

\bibitem[Sharma \& Kiciman(2020)Sharma and Kiciman]{dowhy}
Sharma, A. and Kiciman, E.
\newblock Dowhy: An end-to-end library for causal inference.
\newblock \emph{arXiv preprint arXiv:2011.04216}, 2020.

\bibitem[Sharma et~al.(2022)Sharma, Li, and Jiao]{sharma2022counterfactual}
Sharma, A., Li, H., and Jiao, J.
\newblock The counterfactual-shapley value: Attributing change in system
  metrics.
\newblock In \emph{NeurIPS 2022 Workshop on Causality for Real-world Impact},
  2022.

\bibitem[Shimodaira(2000)]{shimodaira2000improving}
Shimodaira, H.
\newblock Improving predictive inference under covariate shift by weighting the
  log-likelihood function.
\newblock \emph{Journal of statistical planning and inference}, 90\penalty0
  (2):\penalty0 227--244, 2000.

\bibitem[Spirtes et~al.(2000)Spirtes, Glymour, Scheines, and
  Heckerman]{spirtes2000causation}
Spirtes, P., Glymour, C.~N., Scheines, R., and Heckerman, D.
\newblock \emph{Causation, Prediction, and Search}.
\newblock MIT press, 2000.

\bibitem[Sugiyama et~al.(2012)Sugiyama, Suzuki, and
  Kanamori]{sugiyama2012density}
Sugiyama, M., Suzuki, T., and Kanamori, T.
\newblock \emph{Density Ratio Estimation in Machine Learning}.
\newblock Cambridge University Press, 2012.

\bibitem[Tchetgen-Tchetgen \& Shpitser(2012)Tchetgen-Tchetgen and
  Shpitser]{tchetgen2012semiparametric}
Tchetgen-Tchetgen, E.~J. and Shpitser, I.
\newblock Semiparametric theory for causal mediation analysis: Efficiency
  bounds, multiple robustness, and sensitivity analysis.
\newblock \emph{Annals of Statistics}, 40\penalty0 (3):\penalty0 1816, 2012.

\bibitem[Yamamoto(2012)]{yamamoto2012understanding}
Yamamoto, T.
\newblock Understanding the past: Statistical analysis of causal attribution.
\newblock \emph{American Journal of Political Science}, 56\penalty0
  (1):\penalty0 237--256, 2012.

\bibitem[Zhang et~al.(2011)Zhang, Peters, Janzing, and
  Sch{\"o}lkopf]{zhang2011kernel}
Zhang, K., Peters, J., Janzing, D., and Sch{\"o}lkopf, B.
\newblock Kernel-based conditional independence test and application in causal
  discovery.
\newblock In \emph{Proceedings of the Twenty-Seventh Conference on Uncertainty
  in Artificial Intelligence}, pp.\  804--813. AUAI, 2011.

\bibitem[Zhao et~al.(2023)Zhao, Zhang, Zhu, Lu, Dong, Zhang, Xu, Geng, and
  He]{zhao2023conditional}
Zhao, R., Zhang, L., Zhu, S., Lu, Z., Dong, Z., Zhang, C., Xu, J., Geng, Z.,
  and He, Y.
\newblock Conditional counterfactual causal effect for individual attribution.
\newblock In \emph{Uncertainty in Artificial Intelligence}, pp.\  2519--2528.
  PMLR, 2023.

\end{thebibliography}
\bibliographystyle{icml2024}

\newpage
\appendix
\onecolumn
\section{Proofs}
\label{sec:proofs}
\subsection{Proofs of \cref{sec:examp}}
\begin{proof}[Proof of \cref{lem:identexamp}]
We begin by assuming $P^{(1)}_{X} \ll P^{(0)}_{X}$, that is, $P^{(1)}_{X}$ is absolutely continuous with respect to $P^{(0)}_{X}$. We give a formal definition of absolute continuity in \cref{ass:overl} (Weak Overlap), which we will present in the proof of the general identification result in \cref{thm:main}. We also discuss this assumption in \cref{rem:overl} in the main text. By the Radon-Nykodim Theorem \citep{billingsley1995probability}, the Radon-Nykodim (RN) derivative $\alpha(X) := \rm{d}P^{(1)}_{X}/\rm{d}P^{(0)}_{X}(X)$ exists, is almost-surely unique. Furthermore, we assume $\rm{E}_{(0)}[Y^2] < \infty$ and $\rm{E}_{(0)}[\alpha(X)^2] < \infty$, which is a version of \cref{ass:momen} (Moments). Assumptions \ref{ass:overl} and \ref{ass:momen}, combined with the Cauchy-Schwartz inequality, imply that any function $g(X)$ such that $\rm{E}_{(0)}[g(X)^2] < \infty$ will be $P_X^{(1)}$-integrable function $g(X)$, and moreover:
\begin{equation}
    \rm{E}_{(1)}[g(X)] = \int g(x) \rm{d} P^{(1)}_{X}(x) = \int \alpha(x) g(x) \rm{d} P^{(0)}_{X}(x) = \rm{E}_{(0)}[\alpha(X) g(X)] \tag{RN} \label{eq:RN}
\end{equation}

We're now ready to prove our two identification results. Recall the definition of the target parameter:
\[\theta^{\langle 1,0 \rangle} = \int y \, \rm{d}P^{(0)}_{Y \mid X}(y \mid x) \rm{d}P^{(1)}_{X}(x).\]

By \ref{ass:momen}, $\gamma(X) := \rm{E}_{(0)}[Y \mid X]$ exists and $\rm{E}_{(0)}[\gamma(X)^2] < \infty$. By definition of $\gamma(X)$:
\[\rm{E}_{(1)}[\gamma(X)] = \int \gamma(x) \rm{d}P^{(1)}_X(x) = \int \left\lbrace \int y \, \rm{d}P^{(0)}_{Y\mid X}(y \mid x) \right\rbrace \rm{d}P^{(1)}_X(x) = \int y \, \rm{d}P^{(0)}_{Y \mid X}(y \mid x) \rm{d}P^{(1)}_{X}(x),\]
which shows \eqref{eq:reg}. 

On the other hand, by the Law of Iterated Expectations and property \eqref{eq:RN}, \[\rm{E}_{(0)}[\alpha(X)Y] = \rm{E}_{(0)}[\alpha(X)\gamma(X)] = \rm{E}_{(1)}[\gamma(X)],\] which shows \eqref{eq:rew}.
\end{proof}

\begin{proof}[Proof of \cref{lem:exampleDR}]
Consider first the case of $g(X) = \gamma(X)$. In that case, by definition of $\gamma(X)$, $\rm{E}[e(Y, X) \mid X] = \rm{E}[Y - \gamma(X) \mid X] = 0$. By the Law of Iterated Expectations, $\rm{E}_{(0)}[a(X) e(Y, X)] = 0$ for any $a(X)$ such that $\rm{E}_{(0)}[a(X)^2]<\infty$, and so \eqref{eq:DRexamp} and \eqref{eq:reg} give:
\[\rm{E}_{(1)}[\gamma(X)] + \rm{E}_{(0)}[a(X) e(Y, X)] = \rm{E}_{(1)}[\gamma(X)] = \theta^{\langle 1,0 \rangle}.\]

If $a(X) = \alpha(X)$, by \eqref{eq:RN}, $\rm{E}_{(1)}[g(X)] = \rm{E}_{(0)}[\alpha(X) g(X)]$ for any $g(X)$ such that $\rm{E}_{(0)}[g(X)^2]<\infty$, and so \eqref{eq:DRexamp} and \eqref{eq:rew} give:
\[\rm{E}_{(1)}[g(X)] + \rm{E}_{(0)}[\alpha(X) Y] - \rm{E}_{(0)}[\alpha(X) g(X)] = \rm{E}_{(0)}[\alpha(X) Y] = \theta^{\langle 1,0 \rangle}. \qedhere \]
\end{proof}

\subsection{Proofs of \cref{sec:ident}}
\begin{proof}[Proof of \cref{thm:main}]
We assume the following regularity conditions:
\begin{assumption}[Weak Overlap]\label{ass:overl} 
For all $k = 1, \ldots, K$, $P^{(0)}_{X_k \mid \rm{PA}_k}$ and $P^{(1)}_{X_k \mid \rm{PA}_k}$ are mutually absolutely continuous.
\end{assumption}

Recall that, for two measures $\mu, \nu$ on a $\sigma$-algebra $\mathcal{B}$, we say that $\nu \ll \mu$ ($\nu$ is absolutely continuous with respect to $\mu$) whenever $\mu(B) = 0$ implies $\nu(B) = 0$ for $v \in \mathcal{B}$ \citep{billingsley1995probability}. If $\nu \ll \mu$ and $\mu \ll \nu$, we say that $\mu$ and $\nu$ are mutually absolutely continuous. Absolute continuity guarantees that the RN derivatives that go into the weight exist. Intuitively, we are requiring that the distribution of $X_k \mid \rm{PA}_k$ has the same support in samples 0 and 1. This is essential both for regression methods (avoids extrapolating the regression function) and for re-weighting methods (guarantees that we can find observations that will be similar enough to the other sample), as discussed in \cref{rem:overl}.

\begin{assumption}[Moments]\label{ass:momen}
$\rm{E}_{(t)}[h(Y)^2] < \infty$ and $\rm{E}_{(t)}[\alpha_k(X)^2] < \infty$ for all $k = 1, \ldots, K$ and $t \in \{0,1\}$.
\end{assumption}
In particular, \cref{ass:momen} implies $\rm{E}_{(t)}[\gamma_k(Y)^2] < \infty$ for all $k = 1, \ldots, K$ and $t \in \{0,1\}$.

We can now prove the main result. Recall the definition of the general target parameter:
\[\theta^{\bm{c}} = \int h(y) \rm{d}P_{Y \mid \bm X}^{(c_{K+1})}(y \mid \bm{x})\prod_{k=1}^K \rm{d} P^{(c_k)}_{X_k \mid \rm{PA}_k}(x_k \mid \rm{pa}_k).\]

\paragraph{Step 1: Initial Case} We begin by showing that, if $g_1(X_1) = \gamma_1(X_1) := \rm{E}_{(c_2)} [g_{2}(X_1, X_2) \mid X_1]$ or $a_1(X_1) = \alpha_1(X_1) := \rm{d}P^{(c_1)}_{X_1}/\rm{d}P^{(c_2)}_{X_1}(X_1)$ (but not necessarily both), we have:
\[\rm{E}_{(c_1)} [g_1(X_1)] + \rm{E}_{(c_2)}[a_1(X_1)e_1(X_1, X_2)] = \int g_2(x_1, x_2) \rm{d}P^{(c_2)}_{X_2 \mid X_1}(x_2 \mid x_1) \rm{d}P^{(c_1)}_{X_1}(x_1).\]

Consider first the case $g_1(X_1) = \gamma_1(X_1)$. Then, $\rm{E}_{c_2}[e_1(X_1, X_2) \mid X_1] = 0$ and by the Law of Iterated Expectations $\rm{E}_{c_2}[a_1(X_1)e_1(X_1, X_2)] = 0$ for any $a_1(X_1)$ such that $\rm{E}_{(c_2)}[a_1(X_1)^2] < \infty$. The definition of $\gamma_1(X_1)$ implies:
\[\rm{E}_{(c_1)}[\gamma_1(X_1)] = \int \gamma_1(x_1) \rm{d}P^{(1)}_{X_1}(x_1) = \int \left\lbrace \int g_2(x_1, x_2) \rm{d}P^{(c_2)}_{X_2\mid X_1}(x_2 \mid x_1) \right\rbrace \rm{d}P^{(1)}_X(x).\]
as we wanted to show.

On the other hand, if $a_1(X_1) = \alpha_1(X_1)$, by \eqref{eq:RN}, $\rm{E}_{(c_1)}[g_1(X_1)] = \rm{E}_{(c_2)}[\alpha_1(X_1) g_1(X_1)]$ for any $g_1(X)$ such that $\rm{E}_{(c_2)}[g_1(X)^2]<\infty$. Moreover, by the law of iterated expectations and \eqref{eq:RN}, $\rm{E}_{(c_2)}[\alpha_1(X_1) g_2(X_1, X_2)] = \rm{E}_{(c_2)}[\alpha_1(X_1) \gamma_1(X_1)] = \rm{E}_{(c_1)}[\gamma_1(X_1)]$, so
\[\rm{E}_{(c_1)} [g_1(X_1)] + \rm{E}_{(c_2)}[\alpha_1(X_1)g_2(X_1, X_2)] - \rm{E}_{(c_2)}[\alpha_1(X_1)g_1(X_1)] = \int g_2(x_1, x_2) \rm{d}P^{(c_2)}_{X_2 \mid X_1}(x_2 \mid x_1) \rm{d}P^{(c_1)}_{X_1}(x_1),\]
as we wanted to show.

\paragraph{Step 2: Induction} Suppose that for some $2 \leq j \leq K$, it holds that:
\[\rm{E}_{(c_1)} [g_1(X_1)] + \sum_{k=1}^{j-1} \rm{E}_{(c_{k+1})}[a_k(\Bar{\bm{X}}_k) e_k(\Bar{\bm{X}}_{k+1})] = \int g_{j}(\Bar{\bm{X}}_j) \prod_{k=1}^j \rm{d}P^{(c_k)}_{X_k \mid \Bar{\bm{X}}_{k-1}} (x_k \mid \Bar{\bm{X}}_{k-1}).\]
We next show that, if 
\begin{align*}
  g_j(\Bar{\bm{X}}_j) & = \gamma_j(\Bar{\bm{X}}_j) := \rm{E}_{(c_{j+1})} [g_{j+1}(\Bar{\bm{X}}_{j+1}) \mid \Bar{\bm{X}}_j] \\ \text{ or } \quad a_j(\Bar{\bm{X}}_j) & = \alpha_j(\Bar{\bm{X}}_j) := \prod_{k=1}^j \frac{\rm{d}P^{(c_k)}_{X_{k} \mid \Bar{\bm{X}}_{k-1}}}{\rm{d}P^{(c_{j+1})}_{X_{k} \mid \Bar{\bm{X}}_{k-1}}} (X_{k} \mid\Bar{\bm{X}}_{k-1})  
\end{align*}
 (but not necessarily both), we have:
\[\rm{E}_{(c_1)} [g_1(X_1)] + \sum_{k=1}^{j} \rm{E}_{(c_{k+1})}[a_k(\Bar{\bm{X}}_k) e_k(\Bar{\bm{X}}_{k+1})] = \int g_{j+1}(\Bar{\bm{X}}_{j+1}) \prod_{k=1}^{j+1} \rm{d}P^{(c_k)}_{X_k \mid \Bar{\bm{X}}_{k-1}} (x_k \mid \Bar{\bm{X}}_{k-1}).\]

If $g_j(\Bar{\bm{X}}_j) = \gamma_j(\Bar{\bm{X}}_j)$, as before, we have $\rm{E}_{c_{j+1}}[e_j(\Bar{\bm{X}}_{j+1}) \mid \Bar{\bm{X}}_j] = 0$, and by the Law of Iterated Expectations $\rm{E}_{(c_{j+1})}[a_j(\Bar{\bm{X}}_j) e_j(\Bar{\bm{X}}_{j+1})] = 0$ for any $a_j(\Bar{\bm{X}}_j)$ such that $\rm{E}_{(c_{j+1})}[a_j(\Bar{\bm{X}}_j)^2] < \infty$. The inductive hypothesis and the definition of $\gamma_j(\Bar{\bm{X}}_j)$ imply the desired result:
\begin{align*}
\rm{E}_{(c_1)} [g_1(X_1)] & +  \sum_{k=1}^{j-1} \rm{E}_{(c_{k+1})}  [a_k(\Bar{\bm{X}}_k) e_k(\Bar{\bm{X}}_{k+1})] + \rm{E}_{(c_{j+1})}[a_j(\Bar{\bm{X}}_j) e_j(\Bar{\bm{X}}_{j+1})] \\ & = \int \gamma_{j}(\Bar{\bm{X}}_j) \prod_{k=1}^j \rm{d}P^{(c_k)}_{X_k \mid \Bar{\bm{X}}_{k-1}} (x_k \mid \Bar{\bm{X}}_{k-1}) \\ & = \int \left\lbrace \int g_{j+1}(\Bar{\bm{X}}_{j+1}) \rm{d}P^{(c_{j+1})}_{X_{j+1}\mid \Bar{\bm{X}}_j}(x_{j+1} \mid \Bar{\bm{X}}_j) \right\rbrace \prod_{k=1}^j \rm{d}P^{(c_k)}_{X_k \mid \Bar{\bm{X}}_{k-1}} (x_k \mid \Bar{\bm{X}}_{k-1}). \tag{$*$} \label{eq:star}
\end{align*}

Otherwise, suppose that $a_j(\Bar{\bm{X}}_j) = \alpha_j(\Bar{\bm{X}}_j)$. Notice that
$\alpha_j(\Bar{\bm{X}}_j)$ is the RN derivative of 
\[\prod_{k=1}^j \rm{d}P^{(c_{k})}_{X_{k} \mid \Bar{\bm{X}}_{k-1}} \text{ with respect to } \prod_{k=1}^j \rm{d}P^{(c_{j+1})}_{X_{k} \mid \Bar{\bm{X}}_{k-1}}.\]
The first distribution is a counterfactual distribution if $c_k \neq c_{k'}$ for some $k,k' \leq j$, because some causal mechanisms are distributed as in sample 0, and others as in sample 1. The second distribution, however, is the factual joint distribution of $\Bar{\bm{X}}_j$ in sample $c_{j+1}$, by the causal Markov assumption and the order of the variable labels. In this light, $\alpha_j(\Bar{\bm{X}}_j)$ satisfies property \eqref{eq:RN}, and for any $g_j(\Bar{\bm{X}}_j)$ with $\rm{E}_{(c_{j+1})}[g_j(\Bar{\bm{X}}_j)^2] < \infty$, we have 
\[\rm{E}_{(c_{j+1})}[\alpha_j(\Bar{\bm{X}}_j) g_j(\Bar{\bm{X}}_j)] = \int g_j(\Bar{\bm{X}}_j) \prod_{k=1}^j \rm{d}P^{(c_k)}_{X_{k} \mid \Bar{\bm{X}}_{k-1}} (x_k \mid \Bar{\bm{X}}_{k-1}).\]

In particular, by the inductive hypothesis and the Law of Iterated Expectations:
\begin{multline*}
\rm{E}_{(c_1)} [g_1(X_1)] +  \sum_{k=1}^{j-1} \rm{E}_{(c_{k+1})}  [a_k(\Bar{\bm{X}}_k) e_k(\Bar{\bm{X}}_{k+1})] + \rm{E}_{(c_{j+1})}[\alpha_j(\Bar{\bm{X}}_j)g_{j+1}(\Bar{\bm{X}}_{j+1})] -  \rm{E}_{(c_{j+1})}[\alpha_j(\Bar{\bm{X}}_j)g_{j}(\Bar{\bm{X}}_j)] \\ = \rm{E}_{(c_{j+1})}[\alpha_j(\Bar{\bm{X}}_j)\gamma_j(\Bar{\bm{X}}_j)].  
\end{multline*}
Finally, property \eqref{eq:RN} and \eqref{eq:star} give the desired result.

\paragraph{Step 3: Final Step} Before we finish, remember that we adopted the notational convention $g_{K+1}(\Bar{\bm{X}}_{K+1}) := h(Y)$. Therefore, the $K$-th inductive step implies:
\[\rm{E}_{(c_1)} [g_1(X_1)] + \sum_{k=1}^{K} \rm{E}_{(c_{k+1})}[a_k(\Bar{\bm{X}}_k) e_k(\Bar{\bm{X}}_{k+1})] = \int h(y) \rm{d}P^{(c_{K+1})}_{Y \mid \bm X} (y \mid \bm{x}) \prod_{k=1}^{K} \rm{d}P^{(c_k)}_{X_k \mid \Bar{\bm{X}}_{k-1}} (x_k \mid \Bar{\bm{X}}_{k-1}).\]
To conclude the proof, remember that we have labeled the variables so that $\rm{PA}_k$ are part of $\Bar{\bm{X}}_{k-1}$. Because of the causal Markov factorization assumption, $P_{X_k \mid \Bar{\bm{X}}_{k-1}} = P_{X_k \mid \rm{PA}_k}$. In other words, once we condition on its direct causes in the DAG, $X_k$ is independent of the rest of variables that are part of $\Bar{\bm{X}}_{k-1}$.
\end{proof}

\subsection{Proofs of \cref{sec:inference}}
\begin{proof}[Proof of \cref{thm:asymp}]
\citet[Theorem 9]{chernozhukov2023automatic} gives an asymptotic normality and inference result for nested regression functionals.
Here we discuss the assumptions that allow us to apply their results. In the statement of the assumptions, $C < \infty$ denotes a generic positive and finite constant.

\begin{assumption}[Strong Overlap]\label{ass:stroverl} There exists $\epsilon > 0$ such that, for all $k = 1, \ldots, K$,
$\epsilon \leq \Pr(T = 1 \mid \bar{\bm{X}}_k) \leq 1 - \epsilon$ almost surely. Moreover, for all $k = 1, \ldots, K$, $\Vert \hat\alpha_k \Vert_\infty \leq C$.
\end{assumption}
This assumption requires the ``posterior'' probability of observation $\bm{X}_k$ sample 1 to be bounded away from 0 and 1. In other words, it must not be possible to tell which sample any given observation comes from with absolute certainty; thus strengthening \cref{ass:overl}. Notice that, this implies $0 < p < 1$, where $p := \Pr(T=1)$, and that $\alpha_k(\bar{\bm{X}}_k)$ will also be almost surely bounded by Bayes' rule \eqref{eq:bayes}. Since the true weights are bounded, we also impose boundedness of the estimated weights $\hat{\alpha}_k(\bar{\bm{X}}_k)$.

\begin{assumption}[Higher-order Moments]\label{ass:moments2} We assume:
\begin{enumerate}[(i)]
    \item  For some $q > 2$ and all $k = 1, \ldots, K$, $\rm{E}_{(c_k)}[{\gamma}_k(\bar{\bm{X}}_k)^q] \leq C$, $\rm{E}_{(c_k)}[\hat{\gamma}_k(\bar{\bm{X}}_k)^q] \leq C$, $\rm{E}_{(c_{k+1})}[{\gamma}_k(\bar{\bm{X}}_k)^q] \leq C$, and $\rm{E}_{(c_{k+1})}[\hat{\gamma}_k(\bar{\bm{X}}_k)^q] \leq C$.
    \item For all $k = 1, \ldots, K$, $\rm{E}_{(c_{k+1})}[(\gamma_{k+1}(\bar{\bm{X}}_{k+1}) - \gamma_{k}(\bar{\bm{X}}_{k}))^2 \mid \bar{\bm{X}}_{k}] \leq C$.
    \item $V := \frac{1}{1-p}\rm{Var}_{(0)}[\psi_{(0)}(\bar{\bm{X}}_{K+1})] + \frac{1}{p}\rm{Var}_{(1)}[\psi_{(1)}(\bar{\bm{X}}_{K+1})]  \geq c_0 > 0$.
\end{enumerate}
\end{assumption}
The first part guarantees that we can apply the Central Limit Theorem. The second part is a regularity condition that allows for heteroskedasticity in each regression, but requires the conditional variance to be bounded. The third part assumes that the asymptotic variance of the target parameter is bounded away from 0. Recall that we defined $\psi_{(t)}(\bar{\bm{X}}_{K+1})$ to be the part of the estimating equation that depends on data from sample $t$, namely:
\[\psi_{(t)}(\bar{\bm{X}}_{K+1}) = \begin{cases}
\gamma_1(X_{1}) + \sum_{k: c_{k+1} = t}\alpha_k(\Bar{\bm{X}}_{k}) e_k(\Bar{\bm{X}}_{k+1}) \quad & \text{if } c_1 = t, \\ 
\sum_{k: c_{k+1} = t}\alpha_k(\Bar{\bm{X}}_{k}) e_k(\Bar{\bm{X}}_{k+1}) \quad & \text{otherwise}.
\end{cases}\]
\begin{assumption}[Estimation Rates]\label{ass:rates} Let $\Vert f \Vert_{t,2} = (\rm{E}_{(t)}[f(X)^2])^{1/2}$ denote the $\mathcal{L}^2(P^{(t)}_X)$ norm. For $t \in \{0,1\}$ and all $k = 1, \ldots, K$, we assume: \begin{enumerate}[(i)]
    \item  $\Vert \hat{\gamma}_k(\bar{\bm{X}}_k) - \gamma_k(\bar{\bm{X}}_k) \Vert_{t,2} = o_p(1)$ and $\Vert \hat{\alpha}_k(\bar{\bm{X}}_k) - \alpha_k(\bar{\bm{X}}_k) \Vert_{t,2} = o_p(1)$.
    \item $\sqrt{n} \Vert \hat{\gamma}_k(\bar{\bm{X}}_k) - \gamma_k(\bar{\bm{X}}_k) \Vert_{t,2} \Vert \hat{\alpha}_k(\bar{\bm{X}}_k) - \alpha_k(\bar{\bm{X}}_k) \Vert_{t,2} = o_p(1)$ and $\sqrt{n} \Vert \hat{\gamma}_{k+1}(\bar{\bm{X}}_{k+1}) - \gamma_{k+1}(\bar{\bm{X}}_{k+1}) \Vert_{t,2} \Vert \hat{\alpha}_k(\bar{\bm{X}}_k) - \alpha_k(\bar{\bm{X}}_k) \Vert_{t,2} = o_p(1)$.
\end{enumerate}
\end{assumption}
\cref{ass:rates} (i) assumes that both $\hat{\gamma}_k$ and $\hat{\alpha}_k$ are consistent in the mean-square sense for $k = 1, \ldots, K$. \cref{ass:rates} (ii) imposes a trade-off between the rates of convergence of $\hat{\gamma}_k$ and $\hat{\alpha}_k$ that is often encountered in the double/debiased ML literature (see, for example, \citealp{chernozhukov2018double}). As we discuss in the main text (\cref{rem:rates}), the product of the root mean-squared errors of $\hat{\gamma}_k$ and $\hat{\alpha}_k$ has to vanish faster than $n^{-1/2}$. That means that, when we have a high-quality estimator for the regression function, the asymptotic guarantees of the method will still hold even if the weights can only be estimated at a relatively slow rate, and vice versa.
\end{proof}

\section{Structural Equations and Potential Outcomes}\label{sec:structural}
Suppose we are in the setting of the example in \cref{sec:examp}. Here we give a structural equations model (SEM) that implies the DAG in \cref{fig:dagsattr}: 
\begin{align*}
    Y &\leftarrow g_Y(T, X, \varepsilon_Y), \\
    X &\leftarrow g_X(T, \varepsilon_X), \\
    T &\leftarrow \varepsilon_T,
\end{align*}
where $(\varepsilon_Y, \varepsilon_X, \varepsilon_T)$ are mutually independent disturbances of arbitrary dimension, and $\leftarrow$ denotes assignment. Shifting the causal mechanism $P^{(t)}_{X}$ from $t = 0$ to $t = 1$ is equivalent to replacing the structural function $g_X(0, \varepsilon_X)$ with $g_X(1, \varepsilon_X)$. Likewise, shifting the causal mechanism $P^{(t)}_{Y \mid X}$ from $t = 0$ to $t = 1$ is equivalent to replacing the structural function $g_Y(0, X, \varepsilon_Y)$ with $g_Y(1, X, \varepsilon_Y)$.

An intervention that sets $T = t$, or $do(T = t)$, is associated with \emph{potential outcomes} $X(t) := g_X(t, \varepsilon_X)$ and $Y(t) := g_Y(t, X(t), \varepsilon_X)$ \citep{imbens2015causal}. Notice that the SEM does not impose any independence restrictions between $X(0)$, $X(1)$, $Y(0)$, and $Y(1)$. An intervention that sets $T = t$ and $X = x$, or $do(T = t, X = x)$, in turn defines another set of potential outcomes $Y(t, x)$.

\citet[\S4.5]{pearl2009causality} defines the \emph{natural indirect effect} of the transition from $T = 0$ to $T = 1$ as:
\[\rm{IE}_{0,1} := \rm{E}[Y(0, X(1))] - \rm{E}[Y(0)].\]
In words, this is the expected effect of holding $T = 0$ constant, and changing $X$ to whatever value it would have attained if $T$ had been set to 1, $X(1)$.  As long as there are no open backdoor paths relative to $T \rightarrow X$ and $(T, X) \rightarrow Y$,\footnote{More generally, we could allow for pre-treatment covariates $W$ that block those backdoor paths. We discuss this generalization in \cref{sec:covariates}.}
\[\rm{IE}_{0,1} = \theta^{\langle 1, 0 \rangle} - \theta^{\langle 0, 0 \rangle},\]
where $\theta^{\langle 0, 0 \rangle} = \rm{E}_{(0)}[Y]$ can be estimated directly by taking the average of $Y$ in sample 0, and $\theta^{\langle 1, 0 \rangle}$ can be identified as described in \cref{sec:examp}. The mediation formula \citep[equation (4.18)]{pearl2009causality} is capturing the same idea as identification by regression \eqref{eq:reg}.

\citet[\S4.5]{pearl2009causality} also defines a \emph{natural direct effect} of the transition from $T = 0$ to $T = 1$ as:
\[\rm{DE}_{0,1} := \rm{E}[Y(1, X(0))] - \rm{E}[Y(0)] = \theta^{\langle 0, 1 \rangle} - \theta^{\langle 0, 0 \rangle}.\]
This corresponds to the average effect of an intervention that sets $T = 1$ while, at the same time fixing $X$ to the value it would have attained under $T = 0$.

The definitions above imply two ways of decomposing the total effect of $T$ on $Y$, as the difference between a direct effect and the indirect effect of the opposite transition:
\begin{align*}
    \rm{E}[Y(1)] - \rm{E}[Y(0)] & = (\theta^{\langle 1, 1 \rangle} - \theta^{\langle 1, 0 \rangle}) + (\theta^{\langle 1, 0 \rangle} - \theta^{\langle 0, 0 \rangle}) = \rm{IE}_{0,1} - \rm{DE}_{1,0}, \quad \text{but also} \\
    & = (\theta^{\langle 1, 1 \rangle} - \theta^{\langle 0, 1 \rangle}) + (\theta^{\langle 0, 1 \rangle} - \theta^{\langle 0, 0 \rangle}) = \rm{DE}_{0,1} - \rm{IE}_{1,0}.
\end{align*}
Which of the possible decompositions to choose? We suggest two methods in \cref{sec:combin}:  \emph{Shapley values} \citep{shapely1953value, budhathoki2021did}, which average over all possible decompositions, or \emph{along a causal path}, which suggests following an order that respects causal ancestry. We note, however, that our main results are about estimating the $\theta^{\bm c}$ parameters, and so they are valid regardless of the particular decomposition a researcher commits to.

\section{Identification with Pre-Treatment Covariates}\label{sec:covariates}
As discussed in \cref{sec:structural}, under the assumption that $T$ is a randomly assigned treatment, the causal change attribution problem has an interpretation in terms of mediation \citet[\S4.5]{pearl2009causality}. Sometimes, random assignment of $T$ may be implausible, but the researcher may have access to a set of pre-treatment covariates $\bm W$ such that an unconfoundedness assumption holds conditional on $\bm W$. It is still possible to get multiple robustness in that case, with two adjustments: (i) the regression and weights need to control for $\bm W$, and (ii) an additional debiasing term is required, to account for the adjustment on covariates. We explain how to do it in this section.

First, notice that the multiply-robust formula given in \cref{thm:main} also holds conditional on covariates, as formalized in the following result.
\begin{proposition}[Identification with Covariates, Conditional Version]
Let $g_k(\Bar{\bm{X}}_k, \bm W)$, $a_k(\Bar{\bm{X}}_k, \bm W)$ be any candidate functions such that $\rm{E}_{(c_{k+1})}[g_k(\Bar{\bm{X}}_k, \bm W)^2] < \infty$, $\rm{E}_{(c_{k+1})}[a_k(\Bar{\bm{X}}_k, \bm W)^2] < \infty$ for $k = 1, \ldots, K$.
Consider the following estimating equation:
\begin{equation}
\rm{E}_{(c_1)}[g_1(X_1, \bm W) \mid \bm W] + \sum_{k=1}^K \rm{E}_{(c_{k+1})}[a_k(\Bar{\bm{X}}_k, \bm W) e_k(\Bar{\bm{X}}_{k+1}, \bm W) \mid \bm W], \tag{MR2} \label{eq:MR2}
\end{equation}
where $e_k(\Bar{\bm{X}}_{k+1}, \bm W) : = g_{k+1}(\Bar{\bm{X}}_{k+1}, \bm W) - g_{k}(\Bar{\bm{X}}_k, \bm W)$ for $k = 1, \ldots, K$. 

For $k = 1, \ldots, K$ define:
\begin{gather*}
 \gamma_k(\Bar{\bm{X}}_k, \bm W) := \rm{E}_{(c_{k+1})} [g_{k+1}(\Bar{\bm{X}}_{k+1}, \bm W) \mid \Bar{\bm{X}}_k, \bm W] =: \rm{E}[g_{k+1}(\Bar{\bm{X}}_{k+1}, \bm W) \mid \Bar{\bm{X}}_k, T = c_{k+1}, \bm W], \\
\alpha_k(\Bar{\bm{X}}_k, \bm W) := \prod_{j=1}^k \frac{\rm{d}P^{(c_j)}_{X_{j} \mid \Bar{\bm{X}}_{j-1}, \bm W}}{\rm{d}P^{(c_{k+1})}_{X_{j} \mid \Bar{\bm{X}}_{j-1}, \bm W}} (X_{j} \mid \Bar{\bm{X}}_{j-1}, \bm W). 
\end{gather*}

Then \eqref{eq:MR2} is equal to the conditional expectation of $Y$ given $\bm W$ under the counterfactual distribution $P_{Y\mid \bm W}^{(c)}$ if, for every $k = 1,\ldots, K$, either $g_k(\Bar{\bm{X}}_k, \bm W) = \gamma_k(\Bar{\bm{X}}_k, \bm W)$ or $a_k(\Bar{\bm{X}}_k, \bm W) = \alpha_k(\Bar{\bm{X}}_k, \bm W)$, but not necessarily both.   
\end{proposition}
The proof is almost identical to that of \cref{thm:main}, and therefore we omit it here.

An unconditional version of the above can then be obtained by averaging \cref{eq:MR2} over the distribution of the pre-treatment covariates $\bm W$. This identification strategy, however, introduces new nuisance functions, the expectations conditional on $\bm W$. Unless $\bm W$ is discrete with a finite number of values, it will generally not be possible to estimate these at the parametric rate. Hence, below we give an estimating equation that is robust also to the adjustment for covariates. This introduces an additional debiasing term in the equation, the one corresponding to $k = 0$ in the sum.
\begin{proposition}[Identification with Covariates, Unconditional Version]
Let $g_k(\Bar{\bm{X}}_k, \bm W)$, $a_k(\Bar{\bm{X}}_k, \bm W)$ be any candidate functions such that $\rm{E}[g_k(\Bar{\bm{X}}_k, \bm W)^2] < \infty$, $\rm{E}[a_k(\Bar{\bm{X}}_k, \bm W)^2] < \infty$ for $k = 0, \ldots, K$.
Consider the following estimating equation:
\begin{equation}
\rm{E}[g_0(\bm W)] + \sum_{k=0}^K \rm{E}[a_k(\Bar{\bm{X}}_k, \bm W) e_k(\Bar{\bm{X}}_{k+1}, \bm W)], \tag{MR3} \label{eq:MR3}
\end{equation}
where $e_k(\Bar{\bm{X}}_{k+1}, \bm W) : = g_{k+1}(\Bar{\bm{X}}_{k+1}, \bm W) - g_{k}(\Bar{\bm{X}}_k, \bm W)$ for $k = 1, \ldots, K$. 

For $k = 1, \ldots, K$ define:
\begin{gather*}
 \gamma_k(\Bar{\bm{X}}_k, \bm W) := \rm{E}_{(c_{k+1})} [g_{k+1}(\Bar{\bm{X}}_{k+1}, \bm W) \mid \Bar{\bm{X}}_k, \bm W] =: \rm{E}[g_{k+1}(\Bar{\bm{X}}_{k+1}, \bm W) \mid \Bar{\bm{X}}_k, T = c_{k+1}, \bm W], \\
\alpha_k(\Bar{\bm{X}}_k, \bm W) := \frac{\bm 1\{T = c_{k+1}\}}{\Pr(T = c_{k+1} \mid \bm W)} \prod_{j=1}^k \frac{\rm{d}P^{(c_j)}_{X_{j} \mid \Bar{\bm{X}}_{j-1}, \bm W}}{\rm{d}P^{(c_{k+1})}_{X_{j} \mid \Bar{\bm{X}}_{j-1}, \bm W}} (X_{j} \mid \Bar{\bm{X}}_{j-1}, \bm W). 
\end{gather*}

Moreover, let:
\[\gamma_0(\bm W) = \rm{E}[g_1(X_1, \bm W) \mid T = c_1, \bm W], \qquad \alpha_0(\bm W) = \frac{\bm 1\{T = c_{1}\}}{\Pr(T = c_{1} \mid \bm W)}.\]

Then \eqref{eq:MR3} is equal to the (unconditional) expectation of $Y$ under the counterfactual distribution $\int P_{Y \mid \bm W}^{(c)} \mathrm{d}P_{\bm W}$ if, for every $k = 0,\ldots, K$, either $g_k(\Bar{\bm{X}}_k, \bm W) = \gamma_k(\Bar{\bm{X}}_k, \bm W)$ or $a_k(\Bar{\bm{X}}_k, \bm W) = \alpha_k(\Bar{\bm{X}}_k, \bm W)$, but not necessarily both.   
\end{proposition}

Notice that, to adjust for covariates, we need to learn two additional nuisance functions: (i) a final regression step that conditions only on covariates, and (ii) the propensity score, $\Pr(T = 1 \mid \bm W)$. These are the same nuisance functions that appear in other problems in causal inference, such as estimating the ATE under unconfoundedness.

\section{Simplifications to the Identifying Equation}\label{sec:simpl}
In the general formulation of \cref{thm:main}, when there are $K$ explanatory variables we need to estimate a total of $2K$ unknown functions ($K$ regressions and $K$ weights). For certain change vectors $\bm{c}$, however, it is possible to simplify the identifying equation to have fewer unknown functions. In this section, we discuss two settings where simplifications may occur. The first is when $c_k = c_{k+1}$ for some $k$ (i.e., when two consecutive regressions are computed with respect to the same probability distribution), allowing us to apply the Law of Iterated Expectations. The second is when $X_k$ and $X_{k+1}$ are (conditionally) independent.

\begin{remark}[Simplification I: Tower Law]\label{rem:iter}
    Recall the Tower Law for conditional expectations:
    \[\rm{E}[\rm{E}[A \mid B, C] \mid B] = \rm{E}[A \mid  B].\]
    We can apply this property when $c_k = c_{k+1}$ for some $k$ (i.e., when two consecutive regressions are computed with respect to the same probability distribution) because:
    \begin{align*}
        \rm{E}_{(c_k)}[\gamma_k(\Bar{\bm{X}}_k) \mid \Bar{\bm{X}}_{k-1}]
        & =  \rm{E}_{(c_k)}[ \rm{E}_{(c_{k+1})}[\gamma_{k+1}(\Bar{\bm{X}}_{k+1}) \mid \Bar{\bm{X}}_k] \mid \Bar{\bm{X}}_{k-1}] \\
        & = \rm{E}_{(c_k)}[\gamma_{k+1}(\Bar{\bm{X}}_{k+1}) \mid \Bar{\bm{X}}_{k-1}].
    \end{align*}
    Thus, we can skip estimation of $\gamma_k(\Bar{\bm{X}}_k)$, and define $\gamma_{k-1}(\Bar{\bm{X}}_{k-1}) := \rm{E}_{(c_k)}[g_{k+1}(\Bar{\bm{X}}_{k+1}) \mid \Bar{\bm{X}}_{k-1}]$ directly. Because we are not estimating $\gamma_k(\Bar{\bm{X}}_k)$, we can also drop the debiasing term $\rm{E}_{(c_{k+1})}[a_k(\Bar{\bm{X}}_k) e_k(\Bar{\bm{X}}_{k+1})]$, so that we also do not need to estimate the corresponding weight $\alpha_k(\Bar{\bm{X}}_k)$. Finally, we need to adjust the residual in the $(k-1)$-th debiasing term according to the simplification to $e_{k-1}(\Bar{\bm{X}}_{k+1}) = g_{k+1}(\Bar{\bm{X}}_{k+1}) - g_{k-1}(\Bar{\bm{X}}_{k-1})$. 
\end{remark}

\begin{remark}[Simplification II: Conditional Independence]\label{rem:indep}
    Suppose that $A \perp B \mid C$: 
    \[\rm{E}[\rm{E}[g(A, B, C) \mid A, C] \mid C ] = \rm{E}[\rm{E}[g(A, B, C) \mid B, C] \mid C ],\]
    i.e., the order of integration of $A$ and $B$ can be exchanged. We can apply this property when $X_k \perp X_{k+1} \mid \Bar{\bm{X}}_{k-1}$ for some $k$ as:
    \begin{align*}
        \rm{E}_{(c_{k})}[\gamma_k(\Bar{\bm{X}}_k) \mid \Bar{\bm{X}}_{k-1}] & = \rm{E}_{(c_{k})}[ \rm{E}_{(c_{k+1})}[\gamma_{k+1}(\Bar{\bm{X}}_{k+1}) \mid \Bar{\bm{X}}_k] \mid \Bar{\bm{X}}_{k-1}] \\
        & = \rm{E}_{(c_{k+1})}[ \rm{E}_{(c_{k})}[\gamma_{k+1}(\Bar{\bm{X}}_{k+1}) \mid \Bar{\bm{X}}_{k-1}, X_{k+1}] \mid \Bar{\bm{X}}_{k-1}].
    \end{align*}
    Combined with the previous remark, this re-ordering can reduce the number of functions to estimate, as shown in the following examples.
\end{remark}

\begin{figure}[!h]
\centering
\input{figures/dag_simplif}
\caption{DAG with causal factorization $P^{(t)}_{(\bm{X}, Y)} = P^{(t)}_{Y \mid X_2, X_3} P^{(t)}_{X_3 \mid X_1} \allowbreak P^{(t)}_{X_2 \mid X_1} P^{(t)}_{X_1}$}
\label{fig:dag_simpl}  
\end{figure}

\begin{example}
Consider $K = 3$ and the causal Markov factorization depicted in \cref{fig:dag_simpl}. (We omit $T$ from the DAG, with the understanding that the distribution of all the variables depends on $T$.) This factorization implies that $X_2 \perp X_3 \mid X_1$. Suppose we are interested in estimating $\theta^{\bm{c}}$ for $\bm{c} = \langle 1, 0, 1, 0 \rangle$. \cref{rem:indep} implies that we can exchange the order of integration of $X_2$ and $X_3$. Because we're integrating $X_1$ and $X_3$ with respect to the same distribution $P_{(\bm{X}, Y)}^{(1)}$, and $X_2$ and $Y$ with respect to the same distribution $P_{(\bm{X}, Y)}^{(0)}$, \cref{rem:iter} helps us in reducing the number of unknown functions to estimate. In particular, we can use the doubly-robust estimating equation:
\[\rm{E}_{(1)}[g_1(X_1, X_3)] + \rm{E}_{(0)}[a_1(X_1, X_3)\{h(Y) - g_1(X_1, X_3)\}],\]
where the true value of the unknown functions is:
\[\gamma_1(X_1, X_2) = \rm{E}_{(0)}[h(Y) \mid X_1, X_2], \quad \alpha_1(X_1, X_2) = \rm{d}P^{(1)}_{X_1, X_3}/\rm{d}P^{(0)}_{X_1, X_3}(X_1, X_3).\]
Thus, we only need to estimate 2 unknown functions (rather than 6).
\end{example}

\begin{example}
An important example is when all explanatory variables are independent, i.e., the causal Markov factorization is $P_{(\bm{X}, Y)} = P_{Y \mid \bm{X}} \prod_{k=1}^K P_{X_k}$. For a given change vector $\bm{c}$, let $\bm{X}_{\bm{c}} = (X_k : c_k = 1)$ and $\bm{X}_{-\bm{c}} = (X_k : c_k = 0)$. Thanks to Remarks \ref{rem:iter} and \ref{rem:indep} we can give a doubly-robust estimating equation for $\theta^{\bm{c}}$ that only depends only on 2 unknown functions, regardless of what $K$ is. Whenever $c_{K+1} = 0$ (i.e., we want to keep the conditional distribution of $Y \mid \bm{X}$ as in sample 0), the doubly-robust estimating equation will be:
\[\rm{E}_{(1)}[g_1(\bm{X}_{\bm{c}})] + \rm{E}_{(0)}[a_1(\bm{X}_{\bm{c}})\{h(Y) - g_1(\bm{X}_{\bm{c}})\}],\]
where the true value of the unknown functions is:
\[\gamma_1(\bm{X}_{\bm{c}}) = \rm{E}_{(0)}[h(Y) \mid \bm{X}_{\bm{c}}], \quad \alpha_1(\bm{X}_{\bm{c}}) = \rm{d}P^{(1)}_{\bm{X}_{\bm{c}}}/\rm{d}P^{(0)}_{\bm{X}_{\bm{c}}}(\bm{X}_{\bm{c}}).\]
Conversely, if $c_{K+1} = 1$ (i.e., we want to shift the conditional distribution of $Y \mid \bm{X}$ to be as in sample 1), the doubly-robust estimating equation equation will be:
\[\rm{E}_{(0)}[g_1(\bm{X}_{-\bm{c}})] + \rm{E}_{(1)}[a_1(\bm{X}_{-\bm{c}})\{h(Y) - g_1(\bm{X}_{-\bm{c}})\}],\]
where the true value of the unknown functions is:
\[\gamma_1(\bm{X}_{-\bm{c}}) = \rm{E}_{(1)}[h(Y) \mid \bm{X}_{-\bm{c}}], \quad \alpha_1(\bm{X}_{-\bm{c}}) = \rm{d}P^{(0)}_{\bm{X}_{-\bm{c}}}/\rm{d}P^{(1)}_{\bm{X}_{-\bm{c}}}(\bm{X}_{-\bm{c}}). \qedhere\]
\end{example}

\section{Multiplier Bootstrap}\label{sec:boot}
In this section, we describe how to apply the multiplier bootstrap procedure of \citet{belloni2017program} to compute standard errors for $\widehat{\rm{PATH}}_k$ and $\widehat{\rm{SHAP}}_k$.

Let $(\xi^*_i)_{i=1}^n$ be i.i.d. draws, independent of the data, from a random variable with $\rm{E}[\xi] = 0$, $\rm{E}[\xi^2] = 1$ and $\rm{E}[\exp(|\xi|)] < \infty$. Common choices for the distribution of $\xi$ include a re-centered standard exponential distribution (Bayesian bootstrap) and the standard Gaussian distribution (Gaussian multiplier bootstrap). We refer the reader to \citet{belloni2017program} for references.

Define, for each $t \in \{0,1\}$, $\bar{\psi}_{(t)}(\bar{\bm{X}}_{K+1}) = \hat{\psi}_{(t)}(\bar{\bm{X}}_{K+1}) - \widehat{\rm{E}}_{(t)}[\hat{\psi}_{(t)}(\bar{\bm{X}}_{K+1})]$. A multiplier bootstrap draw of $\hat\theta^{\bm c}$ can be computed as:
\[\hat\theta^{\bm{c}, *} = \hat\theta^{\bm c} + \sum_{t\in\{0,1\}} \widehat{\rm{E}}_{(t)}[\xi^* \times \bar{\psi}_{(t)}(\bar{\bm{X}}_{K+1})].\]
For each bootstrap draw of $(\xi^*_i)_{i=1}^n$, the resulting set of $\hat\theta^{\bm{c}, *}$ can be used to compute a bootstrap version of the causal attribution parameters of interest, $\widehat{\rm{PATH}}^*_k$ and $\widehat{\rm{SHAP}}^*_k$. If we repeat this process many times, the standard deviation of $\widehat{\rm{PATH}}^*_k$ and $\widehat{\rm{SHAP}}^*_k$ over the bootstrap distribution will be an asymptotically valid estimator for the standard errors of  $\widehat{\rm{PATH}}_k$ and $\widehat{\rm{SHAP}}_k$, respectively.

\section{Monte-Carlo Simulations}\label{sec:appmc}
\paragraph{Design 1} Our simulation design is based on the causal model of \cref{ex:mediat}.  At each simulation draw, we generate two samples of size $n_0 = n_1 = 1000$, according to the following distributions: 
\begin{align*}
   X_1 & \sim N(1, \sigma_{(t)}^2), \\
   X_2 \mid X_1 & \sim N(\beta_{(t)}X_1, 1), \\
   Y \mid X_1, X_2 & \sim N(X_1 + X_2 + 0.25X_1^2 + \delta_{(t)}X_2^2, 1),
\end{align*}
for $t \in \{0,1\}$. The parameters $\mu_{(t)}:=(\sigma_{(t)}^2, \beta_{(t)}, \delta_{(t)})$ capture the features of the distribution that change between sample 0 and 1, namely the variance of $X_1$, the dependence between $X_2$ and $X_1$, and the dependence between $Y$ and $X_2^2$. In particular, we choose the following values: $\mu_{(0)} = (1, 0.5, 0.25)$ and $\mu_{(1)} = (1.21, 0.2, -0.25)$. This choice of parameters ensures that the distribution of the simulated data satisfies the regularity conditions in Assumptions \ref{ass:overl} and \ref{ass:momen}. The functional of interest will be the mean of $Y$ under different counterfactual distributions. \cref{tab:mcres} presents the results of our simulation (over 1000 draws) for different choices of estimators for the regression and the weights.

\paragraph{Design 2} In a second set of experiments, we consider the effect of increasing $K$. We consider a line DAG, $X_1 \rightarrow X_2 \rightarrow \cdots \rightarrow X_K \rightarrow Y$. In sample $T = 0$, we draw $n_0 = 2,000$ observations according to $X_1 \sim N(0, 1)$ and $X_k \mid X_{k-1} \sim N(0.5X_{k-1}, 0.75)$ for $k = 2, \ldots, K+1$. For each simulation draw, we randomly select 1/10 of the causal mechanisms to shift their mean by 0.2 in sample $T=1$, also with $n_1 = 2,000$. In each simulation draw, we compute the $\widehat{\rm{PATH}}_k$ attribution measure, using LASSO to estimate the regressions, and Logistic Regression with $\ell_1$ penalty (Logit-LASSO) for the weights. In both cases, the penalty is selected by cross-validation. We do this for $K \in \{10, 20, 50, 100\}$.

\begin{table*}\centering
\caption{Monte-Carlo simulation results (over 1000 draws). The top panel of each table shows the Mean Absolute Error (MAE) in the estimates of the counterfactual mean parameters $\theta^{\bm{c}}$ for $\bm{c} \in \{0, 1\}^3$. We omit the results for $\theta^{\langle 0, 0, 0 \rangle}$ and $\theta^{\langle 1, 1, 1 \rangle}$, since these parameters can be estimated as sample means directly from the data. The bottom panel of each table shows the MAE in the estimates of the Shapley values corresponding to each causal mechanism. We compare methods that rely only on regression or re-weighting with our multiply-robust (MR) proposed estimator.}\label{tab:mcres}

\vspace{1em}
\subfloat[Regression and Weights Correctly Specified]{\label{tab:mca}
\resizebox{0.45\textwidth}{!}{\begin{tabular}{rccc}
\toprule
& \multicolumn{3}{c}{Mean Absolute Error $\pm$ std. err.} \\ \cmidrule{2-4}
& Regression & Re-weighting & MR \\
\midrule
$\langle 0, 0, 1 \rangle$ & $0.060 \pm 0.001$ & $0.065 \pm 0.002$ & $0.060 \pm 0.001$ \\
$\langle 0, 1, 0 \rangle$ & $0.059 \pm 0.001$ & $0.065 \pm 0.002$ & $0.059 \pm 0.001$ \\
$\langle 0, 1, 1 \rangle$ & $0.057 \pm 0.001$ & $0.057 \pm 0.001$ & $0.057 \pm 0.001$ \\
$\langle 1, 0, 0 \rangle$ & $0.072 \pm 0.002$ & $0.075 \pm 0.002$ & $0.072 \pm 0.002$ \\
$\langle 1, 0, 1 \rangle$ & $0.063 \pm 0.002$ & $0.077 \pm 0.002$ & $0.063 \pm 0.002$ \\
$\langle 1, 1, 0 \rangle$ & $0.064 \pm 0.002$ & $0.078 \pm 0.002$ & $0.064 \pm 0.002$ \\
\midrule
$\rm{SHAP}_1$ & $0.072 \pm 0.002$ & $0.073 \pm 0.002$ & $0.072 \pm 0.002$ \\
$\rm{SHAP}_2$ & $0.036 \pm 0.001$ & $0.043 \pm 0.001$ & $0.037 \pm 0.001$ \\
$\rm{SHAP}_3$ & $0.040 \pm 0.001$ & $0.046 \pm 0.001$ & $0.041 \pm 0.001$ \\
\bottomrule
\end{tabular}}} \qquad
\subfloat[Weights Incorrectly Specified]{\label{tab:mcb}
\resizebox{0.45\textwidth}{!}{\begin{tabular}{rccc}
\toprule
& \multicolumn{3}{c}{Mean Absolute Error $\pm$ std. err.} \\ \cmidrule{2-4}
& Regression & Re-weighting & MR \\
\midrule
$\langle 0, 0, 1 \rangle$ & $0.060 \pm 0.001$ & $0.059 \pm 0.001$ & $0.060 \pm 0.001$ \\
$\langle 0, 1, 0 \rangle$ & $0.059 \pm 0.001$ & $0.071 \pm 0.002$ & $0.059 \pm 0.001$ \\
$\langle 0, 1, 1 \rangle$ & $0.057 \pm 0.001$ & $0.074 \pm 0.002$ & $0.057 \pm 0.001$ \\
$\langle 1, 0, 0 \rangle$ & $0.072 \pm 0.002$ & $0.085 \pm 0.002$ & $0.072 \pm 0.002$ \\
$\langle 1, 0, 1 \rangle$ & $0.063 \pm 0.002$ & $0.060 \pm 0.001$ & $0.063 \pm 0.002$ \\
$\langle 1, 1, 0 \rangle$ & $0.064 \pm 0.002$ & $0.101 \pm 0.002$ & $0.064 \pm 0.002$ \\
\midrule
$\rm{SHAP}_1$ & $0.072 \pm 0.002$ & $0.083 \pm 0.002$ & $0.072 \pm 0.002$ \\
$\rm{SHAP}_2$ & $0.036 \pm 0.001$ & $0.036 \pm 0.001$ & $0.036 \pm 0.001$ \\
$\rm{SHAP}_3$ & $0.040 \pm 0.001$ & $0.064 \pm 0.001$ & $0.040 \pm 0.001$ \\
\bottomrule
\end{tabular}}}

\vspace{1em}
\subfloat[Regression Incorrectly Specified]{\label{tab:mcc}
\resizebox{0.45\textwidth}{!}{\begin{tabular}{rccc}
\toprule
& \multicolumn{3}{c}{Mean Absolute Error $\pm$ std. err.} \\ \cmidrule{2-4}
& Regression & Re-weighting & MR \\
\midrule
$\langle 0, 0, 1 \rangle$ & $0.130 \pm 0.002$ & $0.065 \pm 0.002$ & $0.062 \pm 0.001$ \\
$\langle 0, 1, 0 \rangle$ & $0.066 \pm 0.002$ & $0.065 \pm 0.002$ & $0.060 \pm 0.001$ \\
$\langle 0, 1, 1 \rangle$ & $0.071 \pm 0.002$ & $0.057 \pm 0.001$ & $0.057 \pm 0.001$ \\
$\langle 1, 0, 0 \rangle$ & $0.089 \pm 0.002$ & $0.075 \pm 0.002$ & $0.073 \pm 0.002$ \\
$\langle 1, 0, 1 \rangle$ & $0.098 \pm 0.002$ & $0.077 \pm 0.002$ & $0.064 \pm 0.002$ \\
$\langle 1, 1, 0 \rangle$ & $0.069 \pm 0.002$ & $0.078 \pm 0.002$ & $0.066 \pm 0.002$ \\
\midrule
$\rm{SHAP}_1$ & $0.084 \pm 0.002$ & $0.073 \pm 0.002$ & $0.073 \pm 0.002$ \\
$\rm{SHAP}_2$ & $0.045 \pm 0.001$ & $0.043 \pm 0.001$ & $0.038 \pm 0.001$ \\
$\rm{SHAP}_3$ & $0.081 \pm 0.002$ & $0.046 \pm 0.001$ & $0.042 \pm 0.001$ \\
\bottomrule
\end{tabular}}} \qquad
\subfloat[Regression and Weights Incorrectly Specified]{\label{tab:mcd}
\resizebox{0.45\textwidth}{!}{\begin{tabular}{rccc}
\toprule
& \multicolumn{3}{c}{Mean Absolute Error $\pm$ std. err.} \\ \cmidrule{2-4}
& Regression & Re-weighting & MR \\
\midrule
$\langle 0, 0, 1 \rangle$ & $0.130 \pm 0.002$ & $0.059 \pm 0.001$ & $0.113 \pm 0.002$ \\
$\langle 0, 1, 0 \rangle$ & $0.066 \pm 0.002$ & $0.071 \pm 0.002$ & $0.076 \pm 0.002$ \\
$\langle 0, 1, 1 \rangle$ & $0.071 \pm 0.002$ & $0.074 \pm 0.002$ & $0.072 \pm 0.002$ \\
$\langle 1, 0, 0 \rangle$ & $0.089 \pm 0.002$ & $0.085 \pm 0.002$ & $0.089 \pm 0.002$ \\
$\langle 1, 0, 1 \rangle$ & $0.098 \pm 0.002$ & $0.060 \pm 0.001$ & $0.084 \pm 0.002$ \\
$\langle 1, 1, 0 \rangle$ & $0.069 \pm 0.002$ & $0.101 \pm 0.002$ & $0.063 \pm 0.001$ \\
\midrule
$\rm{SHAP}_1$ & $0.084 \pm 0.002$ & $0.083 \pm 0.002$ & $0.084 \pm 0.002$ \\
$\rm{SHAP}_2$ & $0.045 \pm 0.001$ & $0.036 \pm 0.001$ & $0.036 \pm 0.001$ \\
$\rm{SHAP}_3$ & $0.081 \pm 0.002$ & $0.064 \pm 0.001$ & $0.063 \pm 0.001$ \\
\bottomrule
\end{tabular}}}

\vspace{1em}
\subfloat[Non-parametric Regression and Weights (NN)]{\label{tab:mce}
\resizebox{0.45\textwidth}{!}{\begin{tabular}{rccc}
\toprule
& \multicolumn{3}{c}{Mean Absolute Error $\pm$ std. err.} \\ \cmidrule{2-4}
& Regression & Re-weighting & MR \\
\midrule
$\langle 0, 0, 1 \rangle$ & $0.071 \pm 0.002$ & $0.068 \pm 0.002$ & $0.061 \pm 0.001$ \\
$\langle 0, 1, 0 \rangle$ & $0.079 \pm 0.002$ & $0.076 \pm 0.002$ & $0.059 \pm 0.001$ \\
$\langle 0, 1, 1 \rangle$ & $0.083 \pm 0.002$ & $0.059 \pm 0.001$ & $0.057 \pm 0.001$ \\
$\langle 1, 0, 0 \rangle$ & $0.101 \pm 0.003$ & $0.076 \pm 0.002$ & $0.073 \pm 0.002$ \\
$\langle 1, 0, 1 \rangle$ & $0.080 \pm 0.002$ & $0.080 \pm 0.002$ & $0.063 \pm 0.002$ \\
$\langle 1, 1, 0 \rangle$ & $0.070 \pm 0.002$ & $0.097 \pm 0.002$ & $0.064 \pm 0.002$ \\
\midrule
$\rm{SHAP}_1$ & $0.080 \pm 0.002$ & $0.068 \pm 0.002$ & $0.072 \pm 0.002$ \\
$\rm{SHAP}_2$ & $0.053 \pm 0.001$ & $0.061 \pm 0.001$ & $0.037 \pm 0.001$ \\
$\rm{SHAP}_3$ & $0.051 \pm 0.001$ & $0.061 \pm 0.001$ & $0.041 \pm 0.001$ \\
\bottomrule
\end{tabular}}} \qquad
\subfloat[Non-parametric Regression and Weights (GBoost)]{\label{tab:mcf}
\resizebox{0.45\textwidth}{!}{\begin{tabular}{rccc}
\toprule
& \multicolumn{3}{c}{Mean Absolute Error $\pm$ std. err.} \\ \cmidrule{2-4}
& Regression & Re-weighting & MR \\
\midrule
$\langle 0, 0, 1 \rangle$ & $0.060 \pm 0.001$ & $0.265 \pm 0.002$ & $0.061 \pm 0.001$ \\
$\langle 0, 1, 0 \rangle$ & $0.062 \pm 0.001$ & $0.080 \pm 0.002$ & $0.062 \pm 0.001$ \\
$\langle 0, 1, 1 \rangle$ & $0.058 \pm 0.001$ & $0.065 \pm 0.002$ & $0.058 \pm 0.001$ \\
$\langle 1, 0, 0 \rangle$ & $0.074 \pm 0.002$ & $0.118 \pm 0.002$ & $0.074 \pm 0.002$ \\
$\langle 1, 0, 1 \rangle$ & $0.064 \pm 0.002$ & $0.280 \pm 0.002$ & $0.064 \pm 0.002$ \\
$\langle 1, 1, 0 \rangle$ & $0.065 \pm 0.002$ & $0.139 \pm 0.002$ & $0.065 \pm 0.002$ \\
\midrule
$\rm{SHAP}_1$ & $0.072 \pm 0.002$ & $0.058 \pm 0.001$ & $0.072 \pm 0.002$ \\
$\rm{SHAP}_2$ & $0.037 \pm 0.001$ & $0.120 \pm 0.002$ & $0.037 \pm 0.001$ \\
$\rm{SHAP}_3$ & $0.041 \pm 0.001$ & $0.079 \pm 0.002$ & $0.041 \pm 0.001$ \\
\bottomrule
\end{tabular}}}
\end{table*}

\begin{table}
    \centering
    \caption{Monte-Carlo simulation results (over 5000 draws), Design 2. The table shows the Average Worst-Case Absolute Error (as defined in the main text) for the $\rm{PATH}_k$ attribution measure.}
    \label{tab:mc2}

    \vspace{1em}
    \begin{tabular}{rccc}
    \toprule
    & \multicolumn{3}{c}{Avg. Worst-Case AE $\pm$ std. err.} \\
    \cmidrule{2-4}
    $K$ & Regression & Re-weighting & MR \\ \midrule
10  & $0.051 \pm 0.001$ & $0.041 \pm 0.001$ & $0.030 \pm 0.001$ \\
20  & $0.052 \pm 0.001$ & $0.042 \pm 0.001$ & $0.030 \pm 0.001$ \\
50  & $0.050 \pm 0.001$ & $0.044 \pm 0.001$ & $0.033 \pm 0.001$ \\
100 & $0.053 \pm 0.001$ & $0.055 \pm 0.002$ & $0.044 \pm 0.001$ \\
\bottomrule
    \end{tabular}
\end{table}

\FloatBarrier
\section{Additional Figures for the Real-World Application}
\begin{figure}[!h]
    \centering
    \subfloat[Female vs. male distribution of education (data from CPS2015). \label{fig:educ}]{\includegraphics[width=0.4\columnwidth]{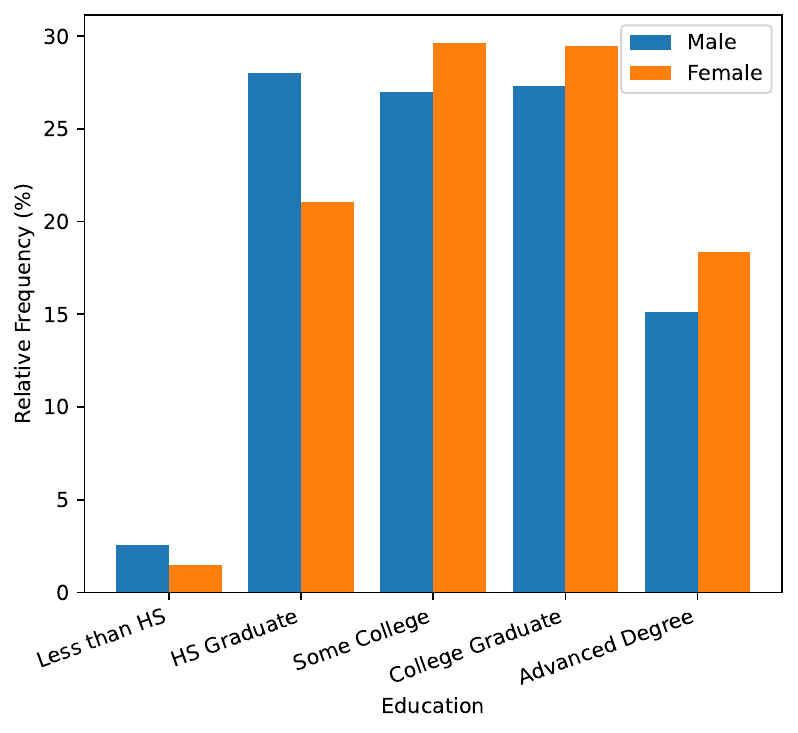}} \qquad
    \subfloat[Female vs. male distribution of occupation conditional on having a college degree (data from CPS2015). \label{fig:occup}]{\includegraphics[width=0.4\columnwidth]{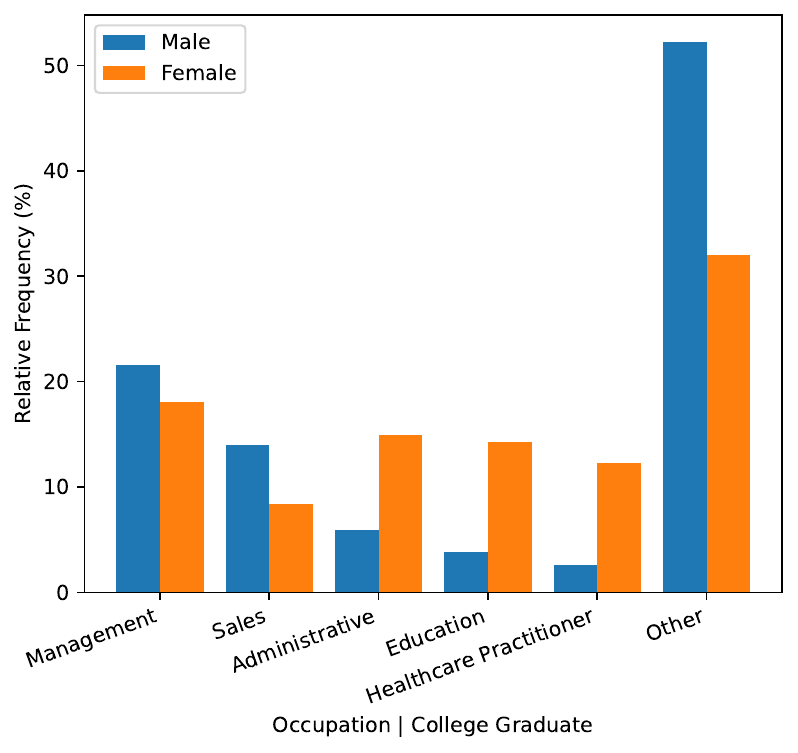}}
    \caption{Distribution of explanatory variables by gender in our gender wage gap application.}
\end{figure}

\end{document}